\documentclass[3p,a4paper,10pt,oneside,onecolumn]{elsarticle}

\usepackage{lineno}
\modulolinenumbers[5]

\usepackage{pifont}
\usepackage{natbib}
\usepackage{geometry}
\usepackage[fleqn]{nccmath}
\usepackage{txfonts}

\usepackage{caption}
\usepackage{booktabs}

\usepackage{amsbsy}
\usepackage{amsfonts}
\usepackage{amsmath}
\usepackage{amssymb}
\usepackage{array}
\usepackage{bbding}
\usepackage{bbm}
\usepackage{bm}
\usepackage{epsfig}
\usepackage{fixmath}
\usepackage[T1]{fontenc}
\usepackage{graphicx}
\usepackage{multirow}
\usepackage{placeins}
\usepackage[svgnames]{xcolor}

\DeclareMathAlphabet{\mathantt}{OT1}{antt}{li}{it}
\DeclareMathAlphabet{\mathpzc}{OT1}{pzc}{m}{it}

\providecommand*{\I}{\mathrm{i}}                           %% imaginary unit i
\newcommand{\tr}{\mathrm{tr}}													%% Trace
\renewcommand{\vec}[1]{\mathbold{#1}}
\newcommand{\nn}{\nonumber}
\renewcommand{\d}{\mathrm{d}}
\newcommand{\ie}{i.\hspace{0.08em}{\nolinebreak}e.$\left.\right.$\hspace{-0.3em}}

\newcommand{\rhs}{r.\hspace{0.08em}{\nolinebreak}h.\hspace{0.08em}{\nolinebreak}s.$\left.\right.$\hspace{-0.3em}~}
\newcommand{\wrt}{w.\hspace{0.08em}{\nolinebreak}r.\hspace{0.08em}{\nolinebreak}t.$\left.\right.$\hspace{-0.3em}~}

\newcommand{\newbok}[3]{\left<\hspace{-0.2ex}{#1}\right|{#2}\left|{#3}\hspace{-0.2ex}\right>}

\newcommand{\Airy}[1]{\left\langle #1 \right\rangle_{\hspace{-0.2em}\mathrm{Ai}}}

\newcommand{\nab}{\boldsymbol{\nabla}}

\newcommand{\nh}{n^{\mathrm{sc}}}
\newcommand{\trace}{\mathrm{tr}}
\newcommand{\EF}{\mu}

\definecolor{ao(english)}{rgb}{0.0, 0.5, 0.0}

\journal{Annals of Physics}

\biboptions{sort&compress}

\begin{document}

\begin{frontmatter}

\title{Airy-averaged gradient corrections\\ for two-dimensional fermion gases}

\author[CQT,CA2DM]{Martin-Isbj\"orn~Trappe\corref{cor1}}
\ead{martin.trappe@quantumlah.org}
\author[CQT,DP]{Yink~Loong~Len}
\ead{yinkloong@quantumlah.org}
\author[CQT,YNC,MAJU]{Hui~Khoon~Ng}
\ead{cqtnhk@nus.edu.sg}
\author[CQT,DP,MAJU]{Berthold-Georg~Englert}
\ead{cqtebg@nus.edu.sg}

\cortext[cor1]{Corresponding author}
\address[CQT]{Centre for Quantum Technologies, National University of Singapore, 3 Science Drive 2, Singapore 117543, Singapore}
\address[CA2DM]{Centre for Advanced 2D Materials and Graphene Research Centre, National University of Singapore, 6 Science Drive 2, Singapore 117546, Singapore}
\address[DP]{Department of Physics, National University of Singapore, 2 Science Drive 3, Singapore 117542, Singapore}
\address[YNC]{Yale-NUS College, 16 College Avenue West, Singapore 138527, Singapore}
\address[MAJU]{MajuLab, CNRS-UNS-NUS-NTU International Joint Unit, UMI 3654, Singapore}

\begin{abstract}
Building on the discussion in PRA \textbf{93}, 042510 (2016), we present a systematic derivation of gradient corrections to the kinetic-energy functional and the one-particle density, in particular for two-dimensional systems. We derive the leading gradient corrections from a semiclassical expansion based on Wigner's phase space formalism and demonstrate that the semiclassical kinetic-energy \textit{density} functional at zero temperature cannot be evaluated unambiguously. In contrast, a \textit{density-potential} functional description that effectively incorporates interactions provides unambiguous gradient corrections. Employing an averaging procedure that involves Airy functions, thereby partially resumming higher-order gradient corrections, we facilitate a smooth transition of the particle density into the classically forbidden region of arbitrary smooth potentials. We find excellent agreement of the semiclassical Airy-averaged particle densities with the exact densities for very low but finite temperatures, illustrated for a Fermi gas with harmonic potential energy. We furthermore provide criteria for the applicability of the semiclassical expansions at low temperatures. Finally, we derive a well-behaved ground-state kinetic-energy functional, which improves on the Thomas-Fermi approximation.
\end{abstract}


\end{frontmatter}

%\linenumbers

\section{\label{Intro}Introduction}
%\gray{Things to be deleted eventually in gray.}
Systems with large particle numbers at low temperatures, where quantum effects become pronounced while classical physics still prevails, are a natural playground for semiclassical expansions. Our objective is to use Wigner's phase-space formulation \cite{Wigner1932,Groenewold1946,Moyal1949} for systematic semiclassical approximations that are potentially useful for a large variety of applications. For example, these may range from semiclassical descriptions of atoms \cite{Berge1988} over spectral properties of matter waves in random potentials \cite{TrDeMu2015} to interacting quantum gases in various trap geometries \cite{KazikBGE2001,Miyakawa2008,FangBerge2011,Bienias2013}. The present work is a sequel to \cite{TrLoNgMuBGE2016}, where the leading gradient correction to the kinetic energy for two-dimensional fermion gases is derived. Here we focus on the gradient corrections to the one-particle density in the realm of density functional theory.

Density functional theory (DFT), formulated in 1964 by Hohenberg and Kohn \cite{HohenbergKohn1964}, has since become a powerful method for coping with many-body problems. DFT is a well-established and widely used tool. Especially, applications in chemistry and condensed matter physics rely on techniques from DFT --- in particular its Kohn-Sham-formulation \cite{KohnSham1965,Dreizler1990} --- with nowadays thousands of publications each year\footnote{Only few studies deal with \emph{developments} of DFT, fewer yet with functionals for two-dimensional systems.} \cite{Burke2012,GrossBurke2015}. Of course, due to the unknown explicit form of the ground-state energy $E[n]$ as a functional of the one-particle density $n$, all applications of DFT have to rely on some physical insight and modeling for setting up an approximate expression for $E[n]$. A myriad of such approximate energy functionals have been formulated. Many of them are refined, tuned, and trained to tackle specific classes of problems; see \cite{Burke2012,Peverati2014,GrossBurke2015,Jones2015} for a selection of recent reviews. These massaged functionals tend to perform poorly on physical systems other than those they are adjusted to.

The wave function-based Kohn-Sham scheme is often employed for simulating mesoscopic systems, albeit restricted to moderate system sizes. It yields fairly accurate predictions in three dimensions, but exhibits a problematic transition to two-dimensional systems \cite{Constantin2008AoP,Raesaenen2010b,Chiodo2012}. Those limitations are thought to be overcome via the computationally highly efficient orbital-free DFT --- provided an explicit form of the noninteracting ground-state kinetic-energy functional $E_{\mathrm{kin}}[n]$ is available. The Thomas-Fermi (TF) model \cite{Thomas1927,Fermi1927}, the predecessor of orbital-free DFT, provides the exact kinetic energy of a noninteracting electron gas with uniform distribution in phase space and serves as a first approximation for inhomogeneous systems. For nonuniform systems, a series of corrections in terms of gradients of the particle density is taken into account to improve upon the TF approximation. Several methods are used for calculating the so-called gradient or quantum corrections as an asymptotic series, formally in powers of $\hbar$, where the TF term is the (semi-)classical limit to order $\mathcal O(\nab^0)$ \footnote{Observing that the TF term also includes $\hbar$, which translates the scales of atomic physics to macroscopic scales of immediate experience, we refrain from the common notation $\mathcal O(\hbar)$. Instead, an appropriate notation is $\mathcal O(\nab)$ since each power of $\hbar$ that originates in the semiclassical expansion is accompanied by one power of the gradient, and the TF term is recovered in the limit of a constant potential, that is, at order $\nab^0$.}; see for example \cite{Kirzhnits1957,GrammaticosVoros1979}.
%\blue{Incorporate somehow: We stick to the phrase ``expanding a physical quantity $Q$ in powers of $\hbar$,'' although $\hbar$ is neither small nor large, but rather translates the scales of atomic physics to macroscopic scales of immediate experience.}

We consider semiclassical expansions of physical observables in terms of Wigner's phase-space function \cite{Wigner1932,Groenewold1946,Moyal1949,Imre1967,Wigner1984,Berge1989}. This approach is well suited for the systematic and explicit calculation of small quantum corrections beyond the classical limit in systems with large particle numbers. Gradient corrections in ${D=3}$ spatial dimensions (3D) are available up to $\mathcal O(\nab^6)$ (see for example \cite{Kirzhnits1957,Hodges1973,Murphy1981}) and implemented in DFT codes.

\begin{table}
\begin{center}
\begin{tabular}{cccc} \toprule
$D$ & $\tau^{\mathrm{TF}}[n]$ & $\tau^{\mathrm{sc}}[n]-\tau^{\mathrm{TF}}[n]$ & Ref.\\ \midrule
1 & $\frac{\pi^2\hbar^2}{24m}\big(n(x)\big)^3$ & $-\frac{\hbar^2}{24m}\frac{\left(\partial_xn(x)^{\phantom{X}\hspace{-1.0ex}}\right)^2}{n(x)}^{\phantom{X}\hspace{-1.8ex}}$ & \cite{Holas1991,Salasnich2007,Koivisto2007AoP}\\ \midrule
\multirow{2}{*}{2} & \multirow{2}{*}{$\frac{\pi\hbar^2}{2m}\big(n(\mathbold{r})\big)^2$} & $\frac{\hbar^2}{24m}\,\delta\big(n(\mathbold{r})\big)\big(\nab n(\mathbold{r})\big)^2$ & \cite{Brack2003,DissvanZyl}\\
 & & 0 & \cite{Holas1991,Shao1993,Salasnich2007,Koivisto2007AoP,Putaja2012}\\ \midrule
3 & $\frac{\hbar^2(3\pi^2)^{5/3}}{10\pi^2 m}\big(n(\mathbold{r})\big)^{5/3}$ & $\frac{\hbar^2}{72m}\frac{\left(\nab n(\mathbold{r})^{\phantom{X}\hspace{-1.0ex}}\right)^2}{n(\mathbold{r})}$ & \cite{Kirzhnits1957,Holas1991,Salasnich2007,Koivisto2007AoP}\\ \bottomrule
\end{tabular}
\end{center}
\caption{\label{IntroTable}Thomas-Fermi approximation $E_{\mathrm{kin}}^{\mathrm{TF}}[n]$ and $\mathcal O(\nab^2)$ gradient corrections $E_{\mathrm{kin}}^{\mathrm{sc}}[n]$ of the kinetic energy $E_{\mathrm{kin}}[n]$ as a functional of the one-particle density $n(\vec r)$ at position $\vec r$. We give the ground-state kinetic-energy densities $\tau^{\mathrm{TF}}[n]$ and the corrections $\tau^{\mathrm{sc}}[n]-\tau^{\mathrm{TF}}[n]$ to order $\nab^2$ as presented in the literature for ${D=1,2}$, and 3 dimensions.}
\end{table}
%Beside the question of accuracy, the known gradient corrections for ${D<3}$ suffer from more fundamental problems.
Surprisingly, comparably little attention has been paid to an unresolved puzzle in orbital-free DFT: The hitherto available quantum corrections beyond the Thomas-Fermi approximation for $E_{\mathrm{kin}}[n]$ are not bounded from below for one-dimensional systems and are, at best, ambiguous in two dimensions. In Table \ref{IntroTable} we cite the TF approximation $E_{\mathrm{kin}}^{\mathrm{TF}}[n]$ of the kinetic energy functional $E_{\mathrm{kin}}[n]$ and the semiclassical approximation $E_{\mathrm{kin}}^{\mathrm{sc}}[n]$ that includes gradient corrections up to $\mathcal O(\nab^2)$ as reported to date. The sign of the gradient correction in 1D is negative. Consequently, the according energy functional is not even bounded from below. As derived in \cite{Salasnich2007,Koivisto2007AoP} by the Kirzhnits method \cite{Kirzhnits1957,Hodges1973}, the gradient corrections for two-dimensional systems vanish to all orders in $\nab$, and one may be led to the wrong conclusion that the TF term constitutes the exact kinetic-energy functional. Vanishing $\mathcal O(\nab^2)$ corrections are also found in \cite{DissvanZyl}, where ${\delta\big(n(\vec r)\big)\big(\nab n(\vec r)\big)^2}$ is argued to be zero for the true physical particle density, which is nonzero everywhere. However, in a perturbation-theoretic evaluation consistent within an approximation up to order $\nab^2$, the leading correction to the energy can be evaluated with the TF density \cite{TrLoNgMuBGE2016}, which is indeed zero at the quantum-classical boundary. At this boundary, however, the gradient of the TF density is ambiguous, and the semiclassical gradient expansion of the \textit{density} functional $E_{\mathrm{kin}}[n]$ given in Table \ref{IntroTable} fails to provide leading-order quantum corrections consistently.

Aside from the introduction of ad hoc parameters to obtain gradient corrections in two dimensions (see for example \cite{vanZyl2013}) other approximations such as the average-density approximation \cite{vanZyl2014} or the semiclassical Wigner-Kirkwood method \cite{DissvanZyl,Brack2003} have been considered. Yet, a systematic semiclassical expansion for the kinetic-energy functional that yields \textit{consistent gradient corrections} without adjustable parameters is currently not available. 
%The purpose of the present work is to unravel and resolve --- at least in part --- the puzzle of inconsistent gradient corrections for the low-dimensional ground-state kinetic-energy functionals and to provide particle densities beyond the TF approximation. Extending the analysis given in \cite{TrLoNgMuBGE2016}, 
We shall see that the trouble with the gradient corrections to $E_{\mathrm{kin}}[n]$ in 2D results from the lack of a one-to-one correspondence between density (to $\mathcal O(\nab^2)$) and effective potential energy. Given that we stick to a semiclassical expansion to obtain consistent gradient corrections of the kinetic energy, we need information about the system beyond what the TF density can provide. This information has to come from the external potential, or, more generally, from the effective potential that incorporates interactions \footnote{Where appropriate for brevity we write `potential' instead of `potential energy', which does not introduce confusion in the context of this paper.}. Therefore, we keep both density and effective potential energy as independent variables of the total energy functional. This flexible formulation yields unambiguous kinetic-energy gradient corrections and provides a convenient tool for obtaining the particle density. The methods presented here are also applicable in 3D and 1D \cite{TrBGE2016}.

In a recent analysis by the authors the leading correction to the kinetic energy beyond the TF approximation in 2D is derived and shown to be nonzero \cite{TrLoNgMuBGE2016}. This correction resolves the long-standing dilemma of vanishing gradient corrections to the kinetic energy. While we demonstrate in the present article that the corresponding gradient-corrected particle density to $\mathcal O(\nab^2)$ is not satisfactory, the unphysical features of the $\mathcal O(\nab^2)$ density at the border between classically allowed and forbidden regions are eliminated via Airy averaging. However, the zero-temperature version of the Airy-averaged densities is unreliable at the extrema of the effective potential energy. This shortcoming is shown to originate in a problematic zero-temperature limit that renders both the $\mathcal O(\nab^2)$ and the Airy-averaged density unphysical. But, at a finite, yet tiny, temperature and in contrast to the $\mathcal O(\nab^2)$ density, the Airy-averaged density is fully satisfactory throughout.

We want to stress that densities of Fermi gases beyond the TF approximation are sought-after quantities by themselves. Although TF densities are frequently used, for example in \cite{Martiyanov2010,Fenech2016,Boettcher2016}, their corrections should be investigated to assess the validity and quality of the TF approximation. Furthermore, a more elaborate theoretical description of the spatial region around the quantum-classical boundary will eventually be required in various applications, e.g., for more precise thermometry \cite{Stewart2006,Lu2012,Aikawa2014}. Improved density profiles also have to be considered in the realm of interacting multi-component Fermi gases, for example for itinerant ferromagnetism, where the inter-species interfaces of repulsive fermion components can crucially depend on gradient corrections \cite{Partridge2006,Du2008,Jo2009,Conduit2009,Zwierlein2011,Ketterle2012,TrGrBrRz2016}.

%A gradient-free semiclassical particle density with remarkable improvement over the TF approximation has recently been found for 1D \cite{Burke2015}.

This article is organized as follows: In Section~\ref{E1} we introduce the energy of an interacting many-body system as a functional of both the one-particle density and the effective one-particle potential. We present a systematic semiclassical expansion obtained from Wigner's phase-space formulation in Section~\ref{Wigner}. In particular, we include an approach which involves an averaging procedure with Airy functions to tackle the ambiguities at the quantum-classical boundary. We investigate the semiclassical expansion of the one-particle density for both finite and zero temperature in Section~\ref{Oneparticledensity}. An illustration of our findings is given in Sections~\ref{Semiclassicalparticledensity} and \ref{ssApplHOAiry} for a two-dimensional harmonic oscillator. The subtleties of the zero-temperature limit for the gradient corrections of the particle density are discussed in Section~\ref{Inadequacy}. We turn to the kinetic-energy functional in Section~\ref{Kineticenergyfunctional} and discuss the ambiguities of the gradient corrections of $E_{\mathrm{kin}}[n]$ for two dimensions as reported in \cite{DissvanZyl}, which are related to the quantum-classical boundary. We show that the semiclassical Airy-averaged energy functionals provide systematic and unambiguous improvement over the TF approximation for the energy and, particularly, the particle density, most noticeable at the boundary
between the classically allowed and forbidden regions.

%All integrals run from $-\infty$ to $\infty$ if not indicated otherwise.

\section{\label{E1} Energy functionals}

According to the Hohenberg-Kohn theorem \cite{HohenbergKohn1964,Dreizler1990} the ground-state energy of a system in an external potential can be expressed as a functional of the one-particle density $n(\vec r)$. Conservation of the particle number $N$ is introduced via the Lagrange multiplier $\mu$, the chemical potential, such that the ground-state energy is the extremum ${E_0=\mathrm{extr}_{\{n,\mu\}}\,E[n,\mu]}$, where
\begin{align}\label{gsEnergy2}
E[n,\mu]&=E_{\mathrm{kin}}[n]+E_{\mathrm{ext}}[n]+E_{\mathrm{int}}[n]+\mu\left(N-\int(\d\vec r)\,n(\vec r)\right)
\end{align}
includes the density functionals of the kinetic energy $E_{\mathrm{kin}}[n]$, the potential energy due to an external single-particle potential energy ${E_{\mathrm{ext}}[n]=\int(\d\vec r)\,V_{\mathrm{ext}}(\vec r)\,n(\vec r)}$, and the interaction energy $E_{\mathrm{int}}[n]$. Eventually, we are interested in the kinetic-energy functional $E_{\mathrm{kin}}[n]$ beyond the Thomas-Fermi approximation. Following the approach in \cite{Berge1988,Berge1992,CinalBerge1993,TrLoNgMuBGE2016}, we introduce the Legendre transform
\begin{align}
E_1[V-\mu]=E_{\mathrm{kin}}[n]+\int(\d\vec r)\,\big(V(\vec r)-\mu\big)\,n(\vec r) \label{defineE1b}
\end{align}
of $E_{\mathrm{kin}}[n]$, where the effective potential (energy) is defined as
\begin{align}\label{deltaEkin}
V(\vec r)=\mu-\frac{\delta E_{\mathrm{kin}}[n]}{\delta n(\vec r)}.
\end{align}
%Quantum corrections of the potential functional $E_1[V-\mu]$ can be obtained straightforwardly from Wigner's phase-space formalism.
%Furthermore, $E_1$ immediately yields the ground-state density.
We write the total energy as a functional of the unconstrained variables $V$, $n$, and $\mu$,
\begin{align}\label{EnergyVnmu}
E[V,n,\mu]&=E_1[V-\mu]-\int(\d\vec r)\,n(\vec r)\,\big(V(\vec r)-V_{\mathrm{ext}}(\vec r)\big)+E_{\mathrm{int}}[n]+\mu N.
\end{align}
%From (\ref{defineE1b}) we get
%\begin{align}
%\delta E_1=\int(\d\vec r)\, n(\vec r)\,\delta(V(\vec r)-\mu) ,\label{defineE1}
%\end{align}
%that is, $E_1$ is a functional of ${(V-\mu)}$ only.
At the ground-state energy, the variations of Eq.~(\ref{EnergyVnmu}) \wrt $V$, $n$, and $\mu$ yield
\begin{align}
n(\vec r)&=\frac{\delta E_1[V-\mu]}{\delta V(\vec r)}\label{nasdef} ,\\
V(\vec r)&=V_{\mathrm{ext}}(\vec r)+\frac{\delta E_{\mathrm{int}}[n]}{\delta n(\vec r)} ,\label{Vdef}
\end{align}
and
\begin{align}
N&=-\frac{\partial E_1[V-\mu]}{\partial\mu} ,\label{defineN}
\end{align}
respectively. While being equivalent to both the density-only and the potential-only functional description \cite{Yang2004,Elliott2008,Gross2009,Elliott2010,Gross2011,Burke2013}, the mixed density-potential functional formulation in Eq.~(\ref{EnergyVnmu}) is more flexible. The coupled Eqs.~(\ref{nasdef})--(\ref{defineN}) allow for the elimination of either the density or the effective potential as variables from $E[V,n,\mu]$ in favor of the other variable \footnote{For instance, Eq.~(\ref{Vdef}) holds for all $V$ and $\mu$ at the ground-state density. If we demand Eq.~(\ref{Vdef}) to hold at all positions $\vec r$, the particle density ceases to be a variable of the total energy functional in Eq.~(\ref{EnergyVnmu}). Then, the ground-state density is rather defined by Eq.~(\ref{Vdef}) (and conveniently obtained by Eq.~(\ref{nasdef})) in terms of the effective potential. In general, all three ground-state variables $n$, $V$, and $\mu$ have to be found self-consistently for obtaining the stationary points of Eq.~(\ref{EnergyVnmu}).}. For a given effective potential energy $V$ the particle density $n$ is obtained from Eq.~(\ref{nasdef}), whereas Eq.~(\ref{Vdef}) yields $V$ for a given $n$. In practice, one solves Eqs.~(\ref{nasdef}) and (\ref{Vdef}) self-consistently for a fixed value of $\mu$ and gets the relation between $\mu$ and $N$ by combining Eqs.~(\ref{nasdef}) and (\ref{defineN}). Approximate particle densities, such as the 1D expression reported in \cite{Burke2015}, can be viewed as approximations of the \rhs of Eq.~(\ref{nasdef}).

In this paper we deal with $E_1[V-\mu]$ and Eq.~(\ref{nasdef}), but do not consider Eq.~(\ref{Vdef}), which is standard fare in DFT: With the interaction accounted for in an effective potential energy, which then differs from the external potential energy according to Eq.~(\ref{Vdef}), we can study an \textit{effectively} noninteracting system. For truly noninteracting systems, Eq.~(\ref{Vdef}) directly yields ${V(\vec r)=V_{\mathrm{ext}}(\vec r)}$ \textit{at the ground-state density}.

Approximations of the kinetic energy and the particle density, respectively, can then be introduced by approximating $E_1[V-\mu]$. Once an expression for ${E_1[V-\mu]}$ is given, the $V$-dependence of the particle density in Eq.~(\ref{nasdef}) may be inverted such that the resulting $n$-dependent effective potential energy yields  $E_{\mathrm{kin}}[n]$ via Eq.~(\ref{defineE1b}), if this is feasible and wanted.

The above reformulation of the total energy proves also valuable because the functional ${E_1[V-\mu]}$ defined in Eq.~(\ref{defineE1b}) can be expressed as
\begin{align}
E_1[V-\mu]=\tr\{\mathcal E_T(H-\mu)\},\label{tracef}
\end{align}
see \cite{Berge1992}, with a function $\mathcal E_T$ ($T$ for temperature) of the single-particle Hamiltonian
\begin{align}\label{singlePartHamil}
H(\vec R,\vec P)=\frac{\vec P^2}{2m}+V(\vec R) ,
\end{align}
where $\vec R$ and $\vec P$ are the single-particle position and momentum vector operators, respectively. Thus, we reduced the problem of expressing the kinetic energy $E_{\mathrm{kin}}$ and particle density $n$ of an interacting many-fermion system to the task of evaluating a single-particle trace of an operator. For the purpose of the present work we are content with the approximate nonrelativistic Hamiltonian in Eq.~(\ref{singlePartHamil}). In particular, we are not accommodating an external magnetic field. The single-particle trace in Eq.~(\ref{tracef}) includes the spin multiplicity.

For interacting systems, ${\mathcal E_T(H-\mu)}$ is a complicated operator, and its explicit form is unknown. But we can approximate it by its expression for noninteracting fermions at ${T=0}$,
\begin{align}\label{fHmu}
\mathcal E_0(H-\mu)=(H-\mu)\,\eta(\mu-H),
\end{align}
see \cite{Berge1992}, which has a very good track record \cite{BGE1984,Berge1992,CinalBerge1993}; it is exact if one includes the kinetic-energy contribution to the correlation energy in the density functional for the interaction energy \footnote{This kinetic energy contribution is the difference between the true kinetic energy of the interacting ground state and the kinetic energy obtained from the ground state of the effectively noninteracting system with the external potential energy replaced by $V(\vec r)$.}, as one would do when employing the formalism of Kohn-Sham orbitals. Approximations of $E_{\mathrm{kin}}$ and $n$, respectively, can be introduced by approximating the trace in Eq.~(\ref{tracef}) with the aid of Eq.~(\ref{fHmu}).
%it is exact if one includes the kinetic-energy contribution to the correlation energy in the density function for the interaction energy, as one would do when employing the formalism of Kohn-Sham orbitals.

We shall develop semiclassical expressions that target low temperatures and investigate the zero-temperature limit since any realistic system is at finite temperature, although very close to its ground state for small enough $T$. Requiring Eq.~(\ref{fHmu}) to be the zero-temperature limit of ${\mathcal E_T(H-\mu)}$ at finite $T$, we choose
\begin{align}\label{fHmuT}
\mathcal E_T(H-\mu)=(-k_{\mathrm{B}}T)\,\mathrm{ln}\left(1+\mathrm{e}^{(\mu-H)/k_{\mathrm{B}}T}\right) ,
\end{align}
which reduces to Eq.~(\ref{fHmu}) for ${T\to0}$. The Boltzmann constant is denoted as $k_{\mathrm{B}}$. We recognize, therefore, that $E_1[V-\mu]$ is the logarithm of the grand-canonical partition function, multiplied by $-k_{\mathrm{B}}T$, of fermions with the single-particle energy of Eq.~(\ref{singlePartHamil}). From Eq.~(\ref{fHmuT}) we shall find in Section~\ref{Wigner} that the single-particle density $n(\vec r)$, derived in accordance with Eq.~(\ref{nasdef}), incorporates the Fermi-Dirac distribution as anticipated. A mere replacement of ${\eta(\mu-H)}$ in Eq.~(\ref{fHmu}) by the Fermi-Dirac distribution would introduce additional total derivatives to $n(\vec r)$. The operator in Eq.~(\ref{fHmuT}) yields exact expressions for $E_1[V-\mu]$ in the case of noninteracting fermions for ${T\ge0}$. %, but it is not unique since arbitrary $V$-independent traceless functions may be added to $\mathcal E_T$ without effect on $E_1[V-\mu]$ and $n(\vec r)$.
Aiming at the quantum corrections of Eq.~(\ref{tracef}), we provide a systematic semiclassical expansion of the trace of arbitrary operators in the following section.

%\green{In accordance with Eq.~(\ref{nasdef}), we then find the Fermi-Dirac distribution
%\begin{align}\label{FermiDirac}
%\eta_T(\mu-H)=\left(1+\mathrm{e}^{(H-\mu)/k_{\mathrm{B}}T}\right)^{-1}
%\end{align}
%for the single-particle density $n(\vec r)$, derived from
%\begin{align}\label{densityfromE1}
%\delta E_1[V-\mu]&=\int(\d\vec r)\, n(\vec r)\,\delta V(r)\nn\\
%&=\tr\big\{\eta_T(\mu-H)\,\delta V(r)\big\}.
%\end{align}
%Since ${\eta_{T\to0}(\mu-H)=\eta_0(\EF -H)=\eta(\EF -H)}$, Eq.~(\ref{densityfromE1}) holds for both finite and zero temperature \footnote{\green{A mere replacement of ${\eta(\mu-H)}$ in Eq.~(\ref{fHmu}) by the Fermi-Dirac distribution would introduce additional total derivatives to $n(\vec r)$. The operator in Eq.~(\ref{fHmuT}) yields exact expressions for $E_1[V-\mu]$ in the case of noninteracting fermions for ${T\ge0}$, but it is not unique since arbitrary $V$-independent traceless functions may be added without effect on $E_1[V-\mu]$ and $n(\vec r)$.}}.}

\section{\label{Wigner}Airy-averaged Wigner transforms}

We choose Wigner's phase-space formulation of quantum mechanics to facilitate a systematic expansion of the trace of an operator in terms of quantum corrections beyond the classical limit. Then, semiclassical expansions of the kinetic energy in Eq.~(\ref{defineE1b}) and the particle density in Eq.~(\ref{nasdef}) are also readily available. The standard Wigner function formalism (see for example \cite{Wigner1932,Groenewold1946,Moyal1949,Imre1967,Wigner1984,Berge1989}) is outlined in the following. Details are provided in the Appendix.

One advantage of the Wigner function formalism lies in its ability to express the trace of any operator $A(\vec R,\vec P)$, depending on position operator $\vec R$ and momentum operator $\vec P$, as the phase-space integral
\begin{align}\label{trH}
\trace\{A(\vec R,\vec P)\}=\int\frac{(\d \vec r)(\d\vec p)}{(2\pi\hbar)^D}\,A_W(\vec r,\vec p)
\end{align}
of its Wigner transform $A_W(\vec r,\vec p)$ over the $D$-dimensional phase-space variables $\vec r$ and $\vec p$. In the mathematical literature, the Wigner transform is known as the Weyl symbol. For brevity we will henceforth not write $\vec r$ and $\vec p$ dependencies, where appropriate.

As an example to be used in the sequel, we consider a system of $N$ spin-1/2 fermions with the single-particle Hamiltonian in Eq.~(\ref{singlePartHamil}) and chemical potential $\mu(T)$. From Eqs.~(\ref{nasdef}) and (\ref{fHmuT}), we then find the single-particle density
\begin{align}\label{density}
n(\vec r)=\frac{2}{(2\pi\hbar)^D}\int(\d\vec p)\,\big[\eta_T(\mu-H)\big]_W(\vec r,\vec p),
\end{align}
where the Fermi-Dirac distribution
\begin{align}\label{FermiDirac}
\eta_T(\mu-H)=\left(1+\mathrm{e}^{(H-\mu)/k_{\mathrm{B}}T}\right)^{-1}
\end{align}
obeys ${\eta_{T\to0}(\mu-H)=\eta_0(\EF -H)=\eta(\EF -H)}$, such that Eq.~(\ref{density}) holds for both finite and zero temperature. We emphasize again that the single-particle Hamiltonian $H$ incorporates interactions via the effective potential energy $V$.

If the Wigner transform $A_W$ is known, Eq.~(\ref{trH}) can be evaluated explicitly. We seek a gradient expansion of the energy functional $E_1[V-\mu]$ with the aid of Eqs.~(\ref{fHmu}) and (\ref{fHmuT}), respectively, for both of which the Wigner transform is not known in general. The Wigner transform of arbitrary operator-valued functions $f( A)$ is approximated by
\begin{align}\label{fAW}
\big[f( A)\big]_W&\cong f(A_W)-\frac{\hbar^2}{16}\big\{A_W\mathrm{\Lambda}^2 A_W\big\}f''(A_W)+\frac{\hbar^2}{24}\big\{A_W\mathrm{\Lambda}  A_W\mathrm{\Lambda}  A_W\big\}f'''(A_W);
\end{align}
see \cite{Wigner1932,GrammaticosVoros1979,VonEiff1991} and \ref{SemiclassicalexpansionofWignertransforms}. Here and in the following `$\cong$' refers to approximate expressions that include all contributions of the exact expression up to (and including) order $\nab^2$. The TF term $f(A_W)$ of $\mathcal O(\nab^0)$ is modified by quantum corrections of $\mathcal O(\nab^2)$. Observe that the powers of $\hbar$ are accompanied by equal powers of the two-sided differential operator of the Poisson bracket
\begin{align}
\mathrm{\Lambda} = \overset{\leftarrow}{\partial_{\vec{r}}}\cdot
 \overset{\rightarrow}{\partial_{\vec{p}}}
  -\overset{\leftarrow}{\partial_{\vec{p}}}\cdot
 \overset{\rightarrow}{\partial_{\vec{r}}} ,\label{Lambda} 
%\underset{\leftarrow}{\partial_{\vec r\phantom{p}\hspace{-1ex}}}\cdot\underset{\rightarrow}{\partial_{\vec p}}-\underset{\leftarrow}{\partial_{\vec p}}\cdot\underset{\rightarrow}{\partial_{\vec r\phantom{p}\hspace{-1ex}}} ,\label{Lambda} 
\end{align}
which acts only on the immediately neighboring functions $A_W$ inside the curly-bracket terms in Eq.~(\ref{fAW}).

The semiclassical expansion in Eq.~(\ref{fAW}) is viable in retrospect if the quantum corrections turn out to be much smaller than the classical part. This cannot be guaranteed in general --- we shall find that Eq.~(\ref{fAW}) is unreliable at the border between the classically allowed and forbidden regions. As a cure, we will make use of an averaging procedure with the Airy function $\mathrm{Ai}(x)$ as a weight. Airy functions in the context of crossing the said border are familiar from WKB connection formulas, and have been used, for instance, in \cite{Baltin1972,Balazs1973,Durand1978} to deal with semiclassical expansions at the classical turning point. Airy-averaged expressions that are similar to the ones introduced here, although not referring to Wigner functions at intermediate steps, have been exploited in refinements of the TF model of atoms; see \cite{BGE1984} and Chapter 4 in \cite{Berge1988}. The Airy-averaged analogue of Eq.~(\ref{fAW}) is derived in \ref{SemiclassicalexpansionofWignertransforms} and reads
\begin{align}\label{fAWAiryMainText}
\big[f(A)\big]_W&\cong\int\d x\, \mathrm{Ai}(x)\left[f\big(\tilde{A}_W\big)-\frac{\hbar^2}{16}\big\{A_W\mathrm{\Lambda}^2 A_W\big\}f''\big(\tilde{A}_W\big)\right],
\end{align}
where
\begin{align}\label{fAWAiryMainTextAWtilde}
\tilde{A}_W=A_W+x\big[(\hbar^2/8)\big\{A_W\mathrm{\Lambda}  A_W\mathrm{\Lambda}  A_W\big\}\big]^{1/3}.
\end{align}

For
\begin{align}\label{specialcase}
A_W(\vec r,\vec p)=H_W(\vec r,\vec p)-\mu=\frac{\vec p^2}{2m}+V(\vec r)-\mu
\end{align}
we show in \ref{SemiclassicalexpansionofWignertransforms} that
\begin{align}
&\int(\d\vec p)\,\big[f( A)\big]_W(\vec r,\vec p)\cong\int(\d\vec p)\int\d x\, \mathrm{Ai}(x)\left[f\big(\tilde{A}_W\big)-\frac{\hbar^2(\nab^2V)}{12m}f''\big(\tilde{A}_W\big)\right] ,\label{trfA4}
\end{align}
where $\tilde{A}_W=\tilde{A}_W(\vec r,\vec p)=H_W(\vec r,\vec p)-\mu-x\, a(\vec r)$, with
\begin{align}\label{axr}
a(\vec r)=\frac{|\hbar\nab V(\vec r)|^{2/3}}{2m^{1/3}}.
\end{align}
The primes denote differentiation with respect to the argument of the function. In contrast to Eq.~(\ref{fAW}), the approximation in Eq.~(\ref{trfA4}) includes higher orders than $\mathcal{O}(\nab^2)$ and is reminiscent of the ``resummed $\hbar$ expansions'' of \cite{Bhaduri1977,Durand1978}, for example. However, the momentum integral of Eq.~(\ref{fAW}) coincides with Eq.~(\ref{trfA4}) up to $\mathcal{O}(\nab^2)$. Conversely, Eq.~(\ref{fAW}) follows from the $\mathcal O(\nab^2)$ truncation of the Airy-averaged Wigner transform in Eq.~(\ref{fAWAiryMainText}) for arbitrary operators $A(\vec R,\vec P)$.

The Airy function provides an exact solution of the Schr\"odinger equation for linear potentials. Hence, we can expect an improved expression for $\big[f(A)\big]_W$ in case of nonlinear potentials which are approximately linear in the vicinity of the quantum-classical boundary; see \cite{Berge1988}.

\section{\label{Oneparticledensity}One-particle density}

The semiclassical Wigner transforms in Eqs.~(\ref{fAW}) and (\ref{fAWAiryMainText}) form the basis for our investigation of the kinetic energy from the potential functional $E_1[V-\mu]$ in Eq.~(\ref{defineE1b}), and thus, for the particle density $n(\vec r)$ in Eq.~(\ref{density}). As an integral of $n(\vec r)$ we can expect the global quantity $E_1[V-\mu]$ to be more regular than $n(\vec r)$ itself. Under specific approximations $E_1[V-\mu]$ may be well-behaved while the local quantity $n(\vec r)$ may become unphysical, \ie, negative or complex \footnote{For instance, TF densities of two-component Fermi gases with contact interaction can become negative or complex, depending on the choices of system parameters. Then the variational equations that determine the unphysical density expressions are invalid in the first place; see \cite{TrGrBrRz2016}.}. Therefore, $n(\vec r)$ can help to discriminate appropriate from inappropriate approximations of $E_1[V-\mu]$.
%The detailed analysis of the one-particle density at zero and finite temperature in the present section serves as a prelude for the study of energy functionals in Section~\ref{Kineticenergyfunctional}.

In Sections~\ref{Semiclassicalparticledensity} and \ref{ssApplHOAiry} we give the semiclassical particle densities in 2D, obtained from Eq.~(\ref{density}) by approximating ${\big[\eta_T(\mu-H)\big]_W}$ with Eqs.~(\ref{fAW}) and (\ref{trfA4}), respectively. From Eq.~(\ref{fAW}) we get the semiclassical particle density ${\nh(\vec r,T)\cong n(\vec r)}$ to $\mathcal O(\nab^2)$, while ${n^{\mathrm{Ai}}(\vec r,T)\cong n(\vec r)}$ follows from Eq.~(\ref{trfA4}). We analyze both $\nh(\vec r,T)$ and $n^{\mathrm{Ai}}(\vec r,T)$ for a 2D harmonic oscillator,
\begin{align}\label{Vfor2DHO}
V(\vec r)=\frac12m\omega^2\vec r^2,
\end{align}
and compare with the exact densities. The harmonic oscillator commonly serves as a benchmark system. If a method fails for the harmonic oscillator, one should consider an improved method before trusting results for more complex systems. In fact, we will encounter an incompatibility between the semiclassical gradient expansion and the zero-temperature limit, investigated in detail in Section~\ref{Inadequacy}.

We emphasize that we choose Eq.~(\ref{Vfor2DHO}) only since exact results are available for noninteracting fermions in an external harmonic potential. The following derivations of densities and energies as functionals of $V$ are no different if interactions are taken into account via Eq.~(\ref{Vdef}). In this context, we also note that the quality of self-consistent solutions of Eqs.~(\ref{nasdef})--(\ref{defineN}) depends equally crucially on the quality of the approximate interaction-energy density functional in Eq.~(\ref{Vdef}) as on the quality of the approximate potential functional Eq.~(\ref{nasdef}). In this work we focus on the latter, eyeing semiclassical developments of orbital-free alternatives to KS-DFT --- very much in the spirit of \cite{Burke2015}, which establishes accurate semiclassical expressions for the particle density in 1D. We derive the r.h.s.~of Eq.~(\ref{nasdef}) from approximations for the Wigner function that is integrated in Eq.~(\ref{density}). Our strategy is therefore to study the quality of the various approximations for the Wigner function in Eq.~(\ref{density}), and thus for the density, without confusing the matter by an admixture of approximations for the interaction energy.

The computational cost in calculating density profiles with our orbital-free approach is independent of the particle number $N$. To underscore the feasibility for large $N$ (e.g. $N\sim10^5$) we choose 316 filled harmonic oscillator shells for the examples in Sections~\ref{Semiclassicalparticledensity} and \ref{ssApplHOAiry}, corresponding to ${N=100172}$ unpolarized spin-$1/2$ fermions. Ground-state energies for various $N$ are presented in Section~\ref{AiryPotentialFunctionalNew}.

\subsection{\label{Semiclassicalparticledensity}Semiclassical particle density to order $\nabla^2$}

Using the approximation given in Eq.~(\ref{fAW}), we arrive at the semiclassical approximation
\begin{align}\label{nFiniteT}
\nh(\vec r,T)&=n^\mathrm{TF}(\vec r,T)+\Delta_{\mathrm{qu}}n(\vec r,T)
\end{align}
of Eq.~(\ref{density}), where
\begin{align}
n^\mathrm{TF}(\vec r,T)&=\frac{mk_{\mathrm{B}}T}{\pi\hbar^2}\mathrm{ln}\left(1+\frac{1}{z}\right),\\
\Delta_{\mathrm{qu}}n(\vec r,T)&=-\frac{\nab^2 U}{12\pi k_{\mathrm{B}}T}\frac{z}{(1+z)^2}+\frac{(\nab U)^2}{24\pi (k_{\mathrm{B}}T)^2}\frac{z^2-z}{(1+z)^3},\\
U&=U(\vec r)=V(\vec r)-\mu,
\end{align}
and the fugacity
\begin{align}\label{z}
z=z(\vec r,T)=\mathrm{e}^{U/k_{\mathrm{B}}T}\ .
\end{align}
Equation~(\ref{nFiniteT}) holds up to $\mathcal O(\nab^2)$ and was also obtained in \cite{vanZyl2011} from a Wigner-Kirkwood expansion.

Eventually targeting the ground-state kinetic-energy functional, we are especially interested in the behavior of the semiclassical expansions at small $T$. Moreover, temperatures well below the Fermi temperature, ${k_{\mathrm{B}}T\ll k_{\mathrm{B}}T_{\mathrm{F}}=\mu^{\mathrm{TF}}}$, are demanded for experiments with ultracold quantum gases and even more so for the description of electron gases in metals and other solid state systems, where typical Fermi temperatures are of the order of several $10^4\,$K \cite{Ashcroft1976}. For the exact density we specify the chemical potential with the help of the Sommerfeld expansion in 2D,
\begin{align}\label{mu}
\frac{\mu^{\mathrm{ex}}(T)}{\mu^{\mathrm{TF}}}\approx 1 -\frac{\pi^2}{6}\left(\frac{k_{\mathrm{B}}T}{\mu^{\mathrm{TF}}}\right)^2-\frac{\pi^4}{72}\left(\frac{k_{\mathrm{B}}T}{\mu^{\mathrm{TF}}}\right)^4,
\end{align}
with terms of order ${(k_{\mathrm{B}}T/\mu^{\mathrm{TF}})^6}$ neglected; see \cite{Ashcroft1976}. The truncated Sommerfeld expansion is a suitable approximation for temperatures ${T\ll T_{\mathrm{F}}}$ \footnote{We use the terms of Eq.~(\ref{mu}) up to $\mathcal O(T^4)$ for our numerical calculations. Within the temperature range considered in this work our numerical results do not change qualitatively if the $T^4$ term of Eq.~(\ref{mu}) is omitted. For $T_1=T_\mathrm{F}/2$ we find the appropriate chemical potential numerically.}. The chemical potential $\mu$ that enters our formalism as a Lagrange multiplier in Eq.~(\ref{gsEnergy2}) has to be determined from the self-consistent solution of Eqs.~(\ref{nasdef})--(\ref{defineN}), i.e., eventually from $N=\int(\d\vec r)\,n(\vec r)$ with a $\mu$-dependent $n(\vec r)$. The numerical value of $\mu$ depends on the chosen approximation of $E_1$ and generally differs from both $\mu^{\mathrm{TF}}$ and $\mu^{\mathrm{ex}}$.

%\begin{table}
%\caption{\label{TableTemperatures}Notations and numerical values of temperatures for 6 ($N=42$; $T_{\mathrm{F}}\approx50\,$nK) and 316 ($N=100172$; $T_{\mathrm{F}}\approx2.4\,\mu$K) filled 2D oscillator shells, respectively, for illustratations of the various expressions of the particle densities in Figs.~\ref{P3finiteT}--\ref{P3AiryFiniteT}.}
%\begin{ruledtabular}
%%\setlength\extrarowheight{0.6em}
%\begin{tabular}{cccc}
% & $T$ [$T_{\mathrm{F}}$] & $T$ (6 shells) & $T$ (316 shells)\\
%\hline\\[-0.6em]
%$T_1$ & $1/2$ &  & $1.2\,\mu$K\\
%$T_2$ & $1/8$ &  & $300\,$nK\\
%$T_3$ & $1/20$ &  & $120\,$nK\\
%$T_4$ & $1/50$ & $1\,$nK & $48\,$nK\\
%$T_5$ & $1/800$ &  & $3\,$nK\\
%$T_6$ & $5\times10^{-4}$ &  & $1.2\,$nK\\
%$T_7$ & $5\times10^{-7}$ &  & $1.2\,$pK\\
%\end{tabular}
%\end{ruledtabular}
%\end{table}

\begin{figure}[htb!]
\begin{center}
\includegraphics[width=0.5\linewidth]{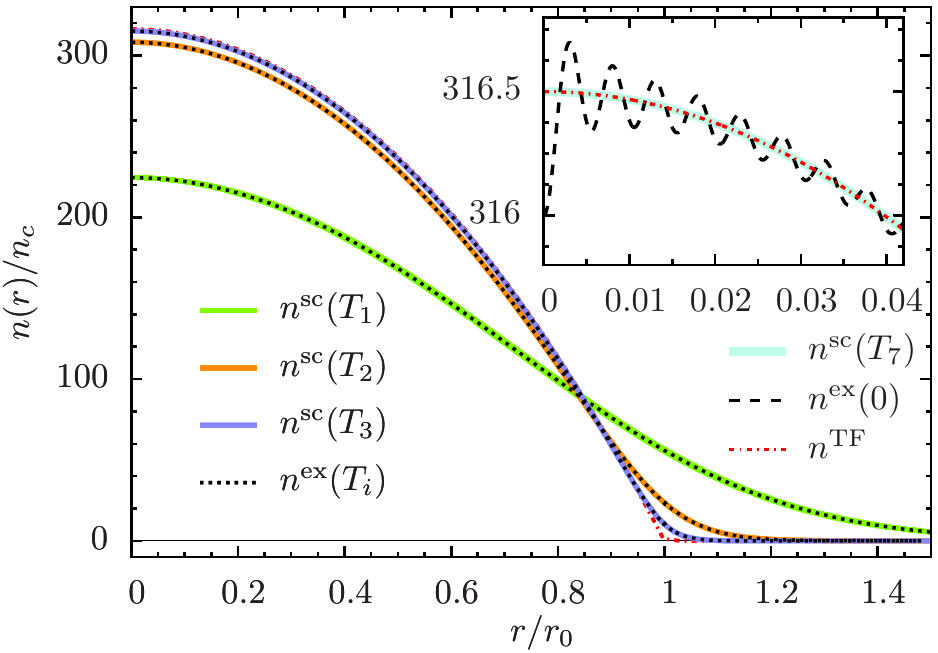}
\caption{\label{P3finiteT}The (isotropic) 2D particle densities for the semiclassical approximation in Eq.~(\ref{nFiniteT}) as functions of the normalized radial coordinate $r/r_0$. Main plot: The chosen temperatures are ${T_1=T_\mathrm{F}/2\approx1.2\,\mu}$K (i.e., $k_{\mathrm{B}}T_1=\hbar\omega\sqrt{N}/2$), ${T_2=T_\mathrm{F}/8\approx300\,}$nK, and ${T_3=T_\mathrm{F}/20\approx120\,}$nK. We observe no visible differences between the semiclassical densities ${\nh(T_i)=\nh(\vec r,T_i)}$ (solid lines) and the exact densities (dotted lines) at the respective temperatures. The inset shows ${\nh(T_7)}$ near the origin, with ${T_7=5\times10^{-7}T_\mathrm{F}\approx1.2\,}$pK, in comparison with the exact ground-state density $n^{\mathrm{ex}}(T=0)$ and the TF density from Eq.~(\ref{TFdensity}). The oscillations of $n^{\mathrm{ex}}(0)$ are not captured by $\nh(T_7)$, which rather accounts for an approximate average over the oscillations and matches $n^{\mathrm{TF}}$ (at the plot resolution).}
\end{center}
\end{figure}

%\begin{figure}[htb!]
%\begin{center}
%\includegraphics[width=0.5\linewidth]{EkinAiry_20160420nAiryFiniteT316ShellsP3bfiniteT}
%\caption{\label{P3bfiniteT}The same particle densities as in Fig.~\ref{P3finiteT}, but for a small range around $\vec r_0$. The exact densities $n^{\mathrm{ex}}(T_i)$ are well resembled by $\nh(T_i)$. For comparison, we show the exact density at ${T=0}$ and the TF density from Eq.~(\ref{TFdensity}), which vanishes for ${r\ge r_0}$.}
%\end{center}
%\end{figure}

\begin{figure}[htb!]
\begin{center}
\includegraphics[width=0.5\linewidth]{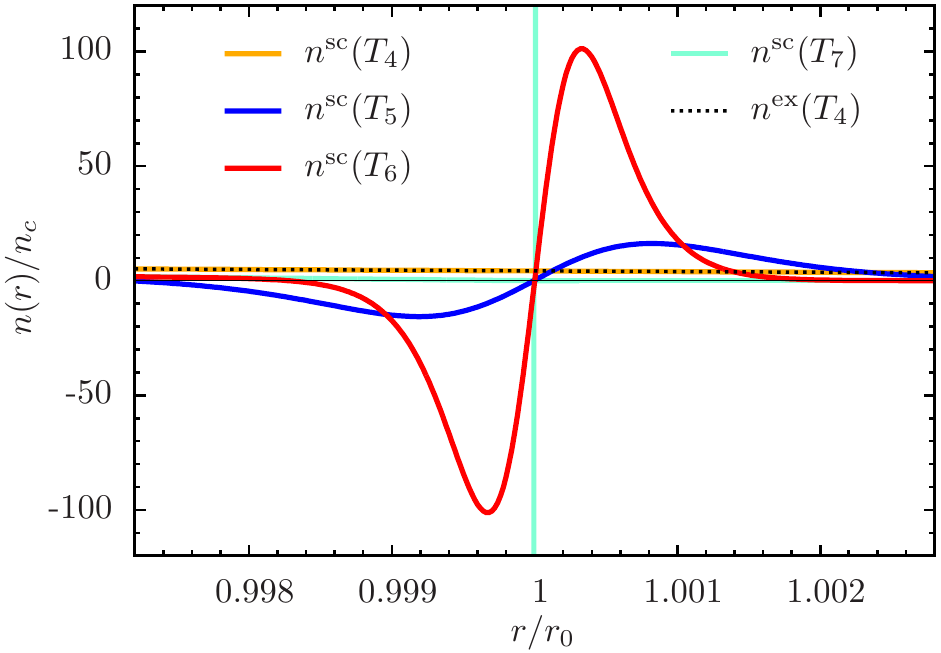}
\caption{\label{P4finiteT}Particle densities $\nh(T_i)$ (solid lines) around $\vec r_0$ for the system parameters as in Fig.~\ref{P3finiteT}, but for the temperatures ${T_4=T_\mathrm{F}/50\approx48\,}$nK, ${T_5=T_\mathrm{F}/800\approx3\,}$nK, and ${T_6=T_\mathrm{F}/2000\approx1.2\,}$nK. The exact density for ${T=T_4}$ is depicted as well. Obviously, $\nh(T)$ turns negative below temperatures of a few nano-Kelvin and exhibits its diverging character, vividly illustrated for ${T=T_7}$.}
\end{center}
\end{figure}

%\begin{figure}[htb!]
%\begin{center}
%\includegraphics[width=0.5\linewidth]{EkinAiry_20160420nAiryFiniteT316ShellsP1P2finiteT}
%\caption{\label{P1finiteT}Plot of $\nh(T)$ around $\vec r_0$ as in Fig.~\ref{P4finiteT}, but for ${T=T_7=5\times10^{-7}T_\mathrm{F}\approx1.2\,}$pK. We also present the exact ground-state density $n^{\mathrm{ex}}(0)$ and the TF density. $\nh(T_7)$ resembles $n^{\mathrm{TF}}$ to a good approximation, except very close to $\vec r_0$, where $n^{\mathrm{TF}}$ has a cusp, while $\nh(T_7)$ clearly exhibits its diverging character. The inset shows the same densities, but near the origin. There the oscillations of $n^{\mathrm{ex}}(0)$ are not captured by $\nh(T_7)$ which rather matches $n^{\mathrm{TF}}$ (at the plot resolution).}
%\end{center}
%\end{figure}

Figures \ref{P3finiteT} and \ref{P4finiteT} show the particle density $\nh(\vec r,T)$ from Eq.~(\ref{nFiniteT}) for various temperatures. The positions $\vec r_0$ given by ${V(\vec r_0)=\EF(T=0)}$ determine the (${D-1}$)-dimensional quantum-classical boundary which we henceforth abbreviate by $\vec r_0$ \footnote{For example, ${|\vec r_0|=r_0\approx32.1\,\mu}$m if we employ 100172 unpolarized $^{40}$K atoms.}. We illustrate the densities in units of ${n_c=m\omega/(\pi\hbar)}$ as a function of ${r/r_0=|\vec r|/r_0}$, with ${\omega=1000\,\mathrm{s}^{-1}}$ \footnote{These normalizations provide density profiles that are independent of the particle mass. Given large $N$ and low $T$, ${n(\vec r=0)/n_c}$ approximately equals the number of filled oscillator shells.}. Figure \ref{P3finiteT} shows excellent agreement between $\nh(\vec r,T)$ and the exact densities $n^{\mathrm{ex}}(\vec r,T)$ taken from \cite{Brack2001,vanZyl2003}. In particular, Fig.~\ref{P3finiteT} does not reveal any inconsistent behavior of $\nh(\vec r,T)$ around $\vec r_0$ at the chosen temperatures. We rather observe a smooth transition into the classically forbidden region of the potential \footnote{The TF density for finite temperature, viz. the leading term of $\nh$ in Eq.~(\ref{nFiniteT}), is indistinguishable (to the eye) from the full expression $\nh$ at $T_1$, $T_2$, $T_3$, and $T_4$.}. However, for lower temperatures we observe unphysical features of $\nh(\vec r,T)$ in a small neighborhood around $\vec r_0$ (see Fig.~\ref{P4finiteT}): $\nh(\vec r,T)$ turns negative, and the magnitude of this anomaly increases as $T$ is lowered. Figure \ref{P3finiteT} includes the TF density
\begin{align}\label{TFdensity}
n^{\mathrm{TF}}(\vec r)=-\frac{m}{\pi\hbar^2}\,U(\vec r)\,\eta\big(-U(\vec r)\big)
\end{align}
at $T=0$ \footnote{The TF densities for ${T>0}$ stay positive across the quantum-classical boundary, in contrast to $\nh(\vec r,T)$ for ${T\lesssim T_5}$.}. For the 2D harmonic oscillator, $n^{\mathrm{TF}}(\vec r)$ integrates to the particle number
\begin{align}\label{particlenumber}
N=\int(\d\vec r)\, n^{\mathrm{TF}}(\vec r)=\left(\frac{m\omega}{2\hbar}\right)^2r_0^4
\end{align}
and drops to zero with a cusp at the radius
\begin{align}\label{R2D}
r_0=\left(\frac{2\hbar}{m\omega}\right)^\frac12 N^\frac14 .
\end{align}
The appropriate chemical potential is determined from Eqs.~(\ref{TFdensity}) and (\ref{particlenumber}) for a given particle number $N$: $\mu=\mu^{\mathrm{TF}}=\frac12 m\omega^2 r_0^2=\hbar\omega\sqrt{N}$. The graphs of $n^{\mathrm{TF}}(\vec r)$ and $\nh(\vec r,T_7)$ are indistinguishable at the resolution of the plot, except very close to $\vec r_0$.

The prime motivation for developing semiclassical approximations of $n(\vec r)$ is to overcome the failure of the TF approximation in the vicinity of $\vec r_0$. There is no a priori reason to believe that those approximations work well deep in the classically forbidden region, where only very small contributions to the total energy $E[n]$ are expected. But the contributions to $E[n]$ from the region around the quantum-classical boundary have to be taken into account for an improvement over the TF approximation. As is obvious from Fig.~\ref{P4finiteT}, $\nh(\vec r,T)$ fails precisely where it should improve matters. Clearly, including only first-order gradient corrections is not sufficient for obtaining a physically reasonable particle density at low enough temperatures. One has to go beyond $\mathcal O(\nab^2)$ to tackle the singular behavior of $\nh(\vec r,T)$ at $\vec r_0$. As we shall demonstrate in the following, the Airy-averaged particle density $n^{\mathrm{Ai}}(\vec r,T)$ provides an excellent description around $\vec r_0$.

%We do not claim that the semiclassical expansion of the density works well deep inside the classically forbidden regions of arbitrary potentials, where only very small contributions to the total energy $E[n]$ are expected. But the contributions to $E[n]$ from the region around the quantum-classical boundary have to be taken into account for an improvement over the TF approximation. There, the semiclassical particle density $\nh(\vec r,T)$, which includes only first-order gradient corrections, fails to provide a physically reasonable description of the particle density at low enough temperatures. One has to go beyond $\mathcal O(\nab^2)$ to tackle the singular behavior of $\nh(\vec r,T)$ at $\vec r_0$. As we will see in the following, the Airy-averaged particle density $n^{\mathrm{Ai}}(\vec r,T)$ provides an excellent description around $\vec r_0$.

\subsection{\label{ssApplHOAiry}Airy-averaged particle density}

In this section we analyze the Airy-averaged particle density ${n^{\mathrm{Ai}}(\vec r,T)\cong n(\vec r)}$. By virtue of the construction of the Airy average it coincides with the semiclassical density in Eq.~(\ref{nFiniteT}) up to $\mathcal O(\nab^2)$; see \ref{SemiclassicalexpansionofWignertransforms}. But it is designed to improve on the description of the particle density in the vicinity of the quantum-classical boundary. No other assessment of the quality of Airy-averaged densities is available in the literature\footnote{This is also true for the three-dimensional precursor in \cite{BGE1984} of the two-dimensional situation studied here. Although Airy-averaged densities of closed Bohr shells were compared with exact densities in \cite{Shakeshaft1985}, no systematic benchmarking was performed. Regarding self-consistent solutions, the somewhat strange features one sees in the radial densities reported in \cite{BGE1984b} are a combined effect of the Airy-averaged gradient corrections of \cite{BGE1984}, the particular way in which strongly-bound electrons are handled, and the approximate treatment of the electron-electron exchange interaction (see also the plots on pages 288--290 in \cite{Berge1988}); the quality of the Airy-averaged gradient corrections themselves cannot be judged in a study of this kind.}.

For the approximation in Eq.~(\ref{trfA4}) the density in Eq.~(\ref{density}) reads
\begin{align}\label{nAirygr0}
n^{\mathrm{Ai}}(\vec r,T)&=\sum_{j=0}^1n_j^{\mathrm{Ai}}(\vec r,T)=\int\d x\,\mathrm{Ai}(x)\,\nu(x,\vec r),
\end{align}
where ${n_j^{\mathrm{Ai}}(\vec r,T)=\int\d x\,\mathrm{Ai}(x)\,\nu_j(\vec r,T)}$,
\begin{align}
\nu(x,\vec r)&=\nu_0(x,\vec r)+\nu_1(x,\vec r),\\
\nu_0(x,\vec r)&=\frac{mk_{\mathrm{B}}T}{\pi\hbar^2}\ln\left(1+\frac{1}{\zeta}\right),\\
\nu_1(x,\vec r)&=-\frac{\nab^2U}{12\pi k_{\mathrm{B}}T}\frac{\zeta}{(1+\zeta)^2},\label{varphi}
\end{align}
and
\begin{align}\label{zeta}
\zeta=\zeta(x,\vec r,T)=\mathrm{e}^{\left[U(\vec r)-x\,a(\vec r)\right]/k_{\mathrm{B}}T} .
\end{align}

We recover the leading (TF) term in Eq.~(\ref{nFiniteT}) from the leading term $n_0^{\mathrm{Ai}}(\vec r,T)$ of Eq.~(\ref{nAirygr0}) if we consider flat potentials (i.e., ${\nab U=0}$, amounting to ${\zeta=z}$) and use ${\int\d x\,\mathrm{Ai}(x)=1}$.

\begin{figure}[htb!]
\begin{center}
\includegraphics[width=0.5\linewidth]{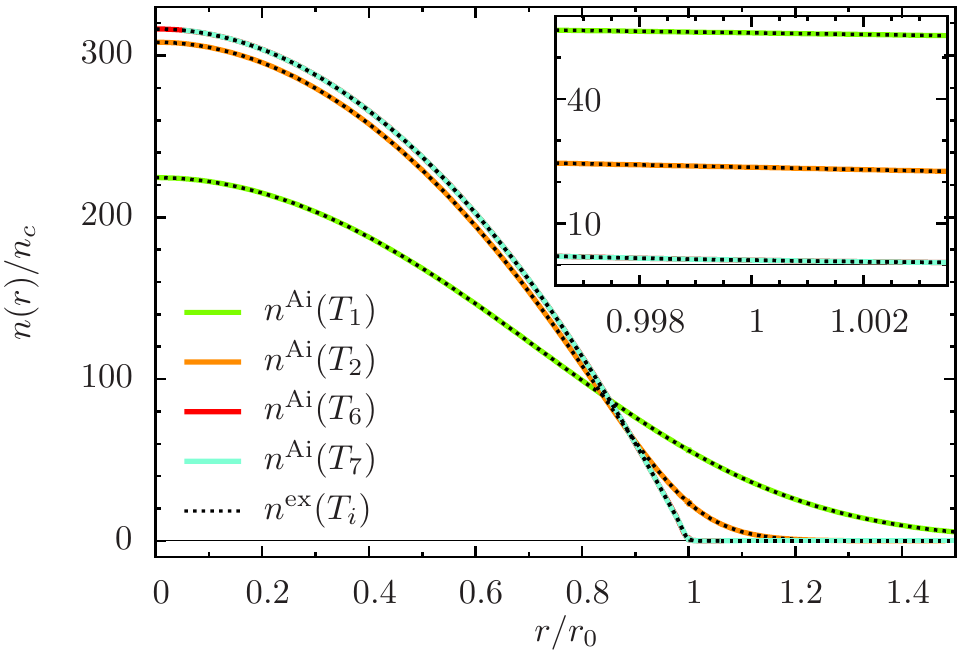}
\caption{\label{nAiryFiniteTFullRange}Plot of the Airy-averaged density given in Eq.~(\ref{nAirygr0}) for a 2D harmonic oscillator at temperatures as noted in Figs.~\ref{P3finiteT} and \ref{P4finiteT}. The densities ${n^{\mathrm{Ai}}(T)=n^{\mathrm{Ai}}(\vec r,T)}$ (colored solid lines) coincide with the exact densities (black dotted lines), and $n^{\mathrm{Ai}}(T_6)$ is indistinguishable from $n^{\mathrm{Ai}}(T_7)$ (at the plot resolution). In particular, the data show an excellent agreement near $\vec r_0$, as illustrated in the inset. For the sake of visibility we depict $n^{\mathrm{Ai}}(T_7)$ only for radii ${r>0.05\,r_0}$.}
\end{center}
\end{figure}
%\begin{figure}[htb!]
%\begin{center}
%\includegraphics[width=0.5\linewidth]{EkinAiry_20160420nAiryFiniteTP4AiryFiniteT}
%\caption{\label{P3AiryFiniteT}The Airy-averaged density as in Fig.~\ref{nAiryFiniteTFullRange} for $T_6$ and $T_7$, shown in the vicinity of $\vec r_0$. There, $n^{\mathrm{Ai}}(T)$ tends to the exact zero-temperature density as the temperature decreases. A larger spatial range is considered in the inset, where the three curves are virtually indistinguishable, while the magnitudes of the densities are much smaller than in the bulk region near the origin.}
%\end{center}
%\end{figure}
\begin{figure}[htb!]
\begin{center}
\includegraphics[width=0.5\linewidth]{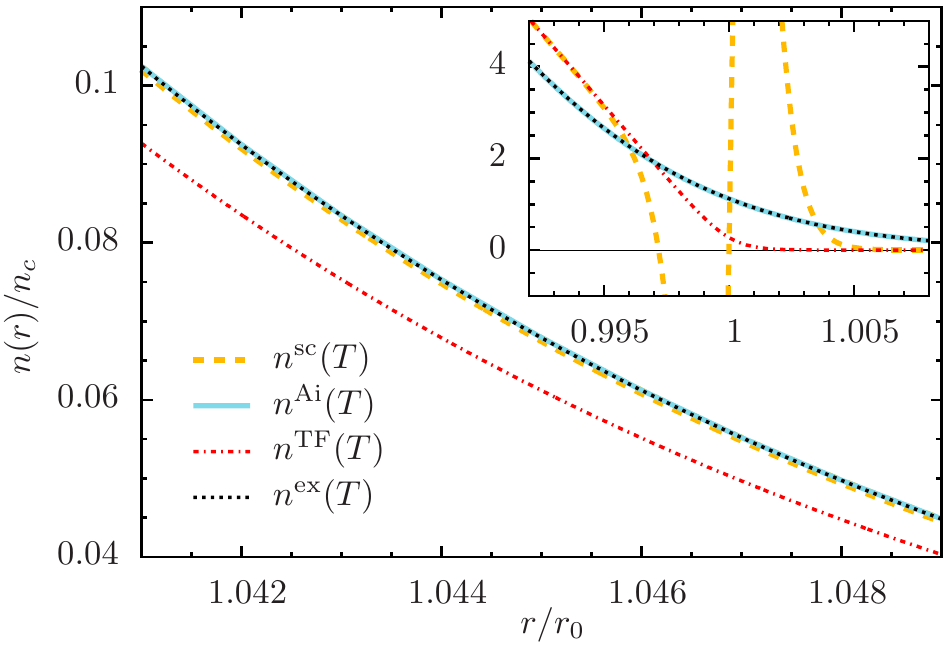}
\caption{\label{nAiryTFhbarExactT5}Comparison of the Airy-averaged density in Eq.~(\ref{nAirygr0}) and the $\mathcal O(\nab^2)$ density in Eq.~(\ref{nFiniteT}) with the exact and the TF densities for 100172 fermions in a 2D harmonic oscillator. The main plot depicts the densities in the classically forbidden region for $T=T_4\approx 48\,$nK, a temperature high enough to ensure $n^{\mathrm{sc}}>0$. The inset shows the same densities, but for $T=T_5\approx 3\,$nK and around the quantum-classical boundary, where the improvement of $n^{\mathrm{Ai}}$ over both $n^{\mathrm{TF}}$ and $n^{\mathrm{sc}}$ is evident.}
\end{center}
\end{figure}

Figures \ref{nAiryFiniteTFullRange} and \ref{nAiryTFhbarExactT5} demonstrate the high quality of the semiclassical approximation $n^{\mathrm{Ai}}(\vec r,T)$ from Eq.~(\ref{nAirygr0}) --- superior to $\nh(\vec r,T)$, cf. Fig.~\ref{P4finiteT}. The densities exhibit a regular behavior across the quantum-classical boundary $\vec r_0$, as opposed to $\nh(\vec r,T)$: There are no differences (to the eye) between $n^{\mathrm{Ai}}(\vec r,T)$ and the exact densities $n^{\mathrm{ex}}(\vec r,T)$ in Fig.~\ref{nAiryFiniteTFullRange}. Both agree especially well around $\vec r_0$, where our numerical data indicate that $n^{\mathrm{Ai}}(\vec r,T)$ tends to the exact zero-temperature density as the temperature decreases. As demonstrated in Fig.~\ref{nAiryTFhbarExactT5} the Airy-averaged density also outperforms the TF density $n^{\mathrm{TF}}(\vec r,T)$, given by the first term of Eq.~(\ref{nFiniteT}), which smoothly reaches beyond $\vec r_0$ at finite $T$.

It is straightforward to show that ${n^{\mathrm{Ai}}(\vec r,T)>0}$ if 
\begin{align}\label{Tcondition}
(k_{\mathrm{B}}T)^2>\frac{\hbar^2}{12m}|\nab^2U(\vec r)|
\end{align}
because Eq.~(\ref{Tcondition}) ensures that ${\nu(-\infty,\vec r)>0}$ and ${\partial_x\nu(x,\vec r)>0}$ for all $x$. Then, as $x$ increases, the growing amplitude and decreasing frequency of the oscillations of $\mathrm{Ai}(x)$ guarantees that the negative contributions to the integral in Eq.~(\ref{nAirygr0}) are overcompensated by the positive ones \footnote{The same line of argument yields ${n_0^{\mathrm{Ai}}(\vec r,T)>0}$ for all $T$.}. For example, the harmonic oscillator employed in this work gives the lower bound $T=\hbar\omega/(k_{\mathrm{B}}\sqrt6)\approx3.12\,$nK, for which ${n^{\mathrm{Ai}}(\vec r,T)>0}$ holds uniformly at all $\vec r$. In fact, ${n^{\mathrm{Ai}}(\vec r,T)}$ is positive even in the pico-Kelvin regime, as indicated in Fig.~\ref{nAiryFiniteTFullRange}. A more in-depth discussion on the positivity of both $n^{\mathrm{sc}}(T)$ and $n^{\mathrm{Ai}}(\vec r,T)$ is provided in Section~\ref{Inadequacy}.

The oscillations of the exact \textit{zero-temperature} density (see the inset of Fig.~\ref{P3finiteT}) are not captured by $n^{\mathrm{Ai}}(\vec r,T)$ and $\nh(\vec r,T)$ even at very low temperatures deep in the classically allowed region, where the semiclassical densities should rather be regarded as an approximate average over the exact oscillations. This is in line with the observation that both $\nh(\vec r,T)$ and $n^{\mathrm{Ai}}(\vec r,T)$ tend to the TF density in Eq.~(\ref{TFdensity}) for ${T\to0}$ if the gradient terms are omitted. 

%\begin{figure}
%\begin{center}
%\includegraphics[width=0.5\linewidth]{EkinAiry_20160420nAiryFiniteTNearOrigin}
%\caption{\label{nAiryFiniteTNearOrigin}Particle densities $n^{\mathrm{Ai}}(T_7)$, $\nh(T_7)$, $n^{\mathrm{TF}}$, and $n^{\mathrm{ex}}(0)$ near the center of the trap. Like the TF density, $n^{\mathrm{Ai}}(T_7)$ as well as $\nh(T_7)$ approximate an average of the oscillations of the exact zero-temperature density. The difference between $n^{\mathrm{ex}}(T_7)$ and $n^{\mathrm{ex}}(0)$ is not resolved at the plot resolution.}
%\end{center}
%\end{figure}

In summary, the finite-temperature Airy-averaged density $n^{\mathrm{Ai}}(\vec r,T)$ demonstrates a significant improvement over $n^{\mathrm{sc}}(\vec r,T)$, especially at very low temperatures, while both are equivalent up to $\mathcal O(\nab^2)$. For the example of the harmonic-oscillator potential we find that the gradient corrections beyond $\mathcal O(\nab^2)$ provided through $n^{\mathrm{Ai}}(\vec r,T)$ are crucial --- and sufficient --- for describing the transition of the particle densities into the classically forbidden region. This observation holds for very low temperatures and even for ${T=0}$ as we will show in the next section.

\subsection{\label{Inadequacy}Zero-temperature limit and regions of validity}

In the previous sections, we focused on the finite-temperature particle densities. Eventually, we are also interested in ground-state expressions of quantities like $E_1[U]$, from which the particle densities are derived via Eq.~(\ref{nasdef}). In the present section we therefore analyze the zero-temperature limit of the semiclassical particle densities $\nh(\vec r,{T})$ and $n^{\mathrm{Ai}}(\vec r,{T})$ from Eqs.~(\ref{nFiniteT}) and (\ref{nAirygr0}), respectively. While $\nh(\vec r,T)$ becomes negative and singular close to $\vec r_0$ as $T$ decreases, $n^{\mathrm{Ai}}(\vec r,T)$ turns out to be well-behaved across $\vec r_0$ in the limit ${T\to0}$ and stays positive except close to potential minima, where $n^{\mathrm{Ai}}(\vec r,T)$ can become negative at very low temperatures. In the following we investigate these anomalies, which can be linked to a troublesome zero-temperature limit of the semiclassical gradient expansion \footnote{In contrast, the semiclassical energies will turn out to be physically well-defined despite an equally ill-defined zero-temperature limit.}, and discuss the parameters for which $n^{\mathrm{Ai}}(\vec r,{T})$ and $\nh(\vec r,{T})$ constitute valid, viz.~positive, particle densities.

%As $T$ goes to zero, the chemical potential $\mu(T)$ tends to the Fermi energy $\EF $. For the other $T$-dependent quantities of $n^{\mathrm{Ai}}(\vec r,T)$ in (\ref{nAirygr0}) we obtain
%\begin{align}
%k_{\mathrm{B}}T\ln\left(1+\frac{1}{\zeta}\right)\longrightarrow\left\{\begin{array}{cl}\hspace{-0.5em}  -\big(V(\vec r)-\mu-\red{x\,a(\vec r)}\big) &\hspace{-0.5em}  ,\,y<x \\ \hspace{-0.5em}  0 &\hspace{-0.5em} ,\,y\ge x\end{array}\right.\hspace{-0.5em} \label{kTlnZeta}
%\end{align}
%and
%\begin{align}
%\frac{\zeta}{k_{\mathrm{B}}T(1+\zeta)^2}=\frac{1}{4k_{\mathrm{B}}T}\frac{1}{\cosh^2\left[(y-x)\frac{t_c}{T}\right]}\longrightarrow\left\{\begin{array}{cl} 0 &\hspace{-0.8ex} ,\; x\not=y \\ %\infty &\hspace{-0.8ex} ,\; x=y\end{array}\right.\label{delta?},
%\end{align}
%where
As $T$ decreases, the crucial feature of the gradient corrections given in Eq.~(\ref{varphi}),
\begin{align}
\nu_1(x,\vec r)&=-\frac{\nab^2U}{48\pi k_{\mathrm{B}}T}\frac{1}{\cosh^2\left[(y-x)\frac{t_c}{T}\right]}=-\frac{\nab^2U}{12\pi}\,K_y(x,T),\label{Ky}
\end{align}
where
\begin{align}\label{R}
k_{\mathrm{B}}\,t_c=\frac{a(\vec r)}{2}=\frac{|\hbar\nab U|^{2/3}}{4m^{1/3}}
\end{align}
and
\begin{align}\label{yr}
y=y(\vec r)=\frac{U(\vec r)}{a(\vec r)}=U\frac{2m^{1/3}}{|\hbar\nab U|^{2/3}} ,
\end{align}
is the increasingly sharp peak of $K_y(x,T)$ at ${x=y}$ with half-width at half-maximum
\begin{align}\label{Deltax}
\Delta x=T\,\mathrm{cosh}^{-1}(\sqrt 2)/t_c.
\end{align}
Indeed, we recognize, with $\zeta$ defined in Eq.~(\ref{zeta}), the $\delta$ function
\begin{align}\label{deltaxy}
\delta(x-y)&=\left.\frac{2t_c\,\zeta}{T(1+\zeta)^2}\right|_{T\to0}=2k_{\mathrm{B}}\,t_c\,K_y(x,{T\to0})\\
&=\left.\frac{t_c}{2T\,\cosh^2\left[(y-x)\frac{t_c}{T}\right]}\right|_{T\to0}=\partial_x\tanh\left[(y-x)\frac{t_c}{T}\right]_{T\to0}\label{delta?3}
\end{align}
for ${T\to0}$ under the $x$-integral. We therefore obtain
\begin{align}\label{nAiryTto0}
&n^{\mathrm{Ai}}(\vec r,{T\to0})=-\frac{m}{\pi\hbar^2}\,U\,\mathcal A(y)-\frac{m^{2/3}}{2\pi\hbar^2}|\hbar\nab U|^{2/3}\mathrm{Ai}'(y)-\frac{m^{1/3}}{6\pi|\hbar\nab U|^{2/3}}\nab^2U\mathrm{Ai}(y) ,
\end{align}
for ${\nab U(\vec r)\not=0}$. Here,
\begin{align}\label{mathcalA}
\mathcal A(y)=\int_y^\infty\d z\,\mathrm{Ai}(z)
\end{align}
is the antiderivative of $-\mathrm{Ai}(y)$.

At the points where ${\nab U(\vec r)=0}$ we find, following the argument after Eq.~(\ref{zeta}),% $\zeta(x,\vec r,T)=z(\vec r,T)$, see Eqs.~(\ref{z}) and (\ref{zeta}). Then, the Airy-averaged density in Eq.~(\ref{nAirygr0}) simplifies since the integration reduces to the familiar integral over $\mathrm{Ai}(x)$, such that
\begin{align}\label{nAiryTto0V0}
n^{\mathrm{Ai}}(\vec r,{T\to0})&=n^{\mathrm{TF}}(\vec r)-\frac{\nab^2U(\vec r)}{12 \pi}\,\delta\big(U(\vec r)\big)
\end{align}
at extrema of $U(\vec r)$. Both Eq.~(\ref{nAiryTto0}) and Eq.~(\ref{nAiryTto0V0}) are also obtained when calculating $n^{\mathrm{Ai}}(\vec r,{T=0})$ directly, that is, when replacing the Fermi-Dirac distribution ${\eta_T(\mu-H)}$ in Eq.~(\ref{density}) by the step function ${\eta(\EF-H)}$. The formal expression of the TF density in Eq.~(\ref{TFdensity}) is reproduced by Eq.~(\ref{nAiryTto0V0}) at potential extrema where $U\not=0$. However, one has to calculate the appropriate chemical potential ${\mu=\mu^{\mathrm{Ai}}}$ for Eqs.~(\ref{nAiryTto0}) and (\ref{nAiryTto0V0}) from
\begin{align}\label{getmuAi}
N=\int(\d\vec r)\,n^{\mathrm{Ai}}(\vec r,T\to0,\mu^{\mathrm{Ai}}), 
\end{align}
while the TF approximation requires
\begin{align}
N=\int(\d\vec r)\,n^{\mathrm{TF}}(\vec r,T\to0,\mu^{\mathrm{TF}}) 
\end{align}
with ${\mu^{\mathrm{Ai}}\not=\mu^{\mathrm{TF}}}$ in general.

\begin{figure}
\begin{center}
\includegraphics[width=0.5\linewidth]{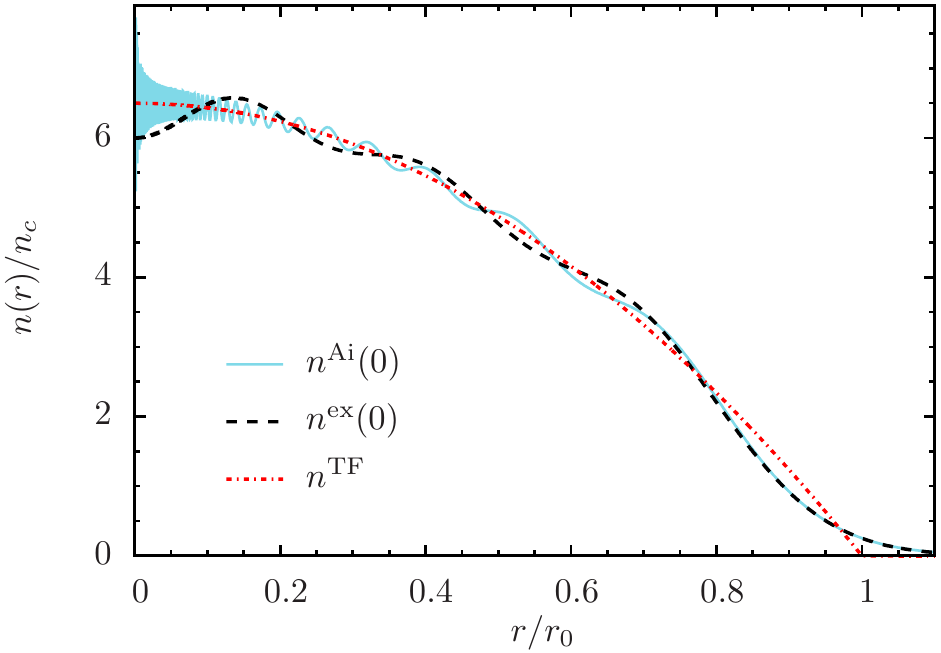}
\caption{\label{nAiryZeroTAiryZeroTFullRange}Zero-temperature particle densities for the isotropic 2D harmonic oscillator with six filled shells, corresponding to 42 fermions. The TF radius is ${r_0\approx4.6\,\mu}$m. We show the Airy-averaged density $n^{\mathrm{Ai}}(0)$, together with the exact density $n^{\mathrm{ex}}(0)$ and the TF density $n^{\mathrm{TF}}$.}
\end{center}
\end{figure}

In Fig.~\ref{nAiryZeroTAiryZeroTFullRange} we depict $n^{\mathrm{Ai}}(\vec r,0)$ from Eq.~(\ref{nAiryTto0}) for the 2D harmonic oscillator, together with the exact ground-state density and the TF density. In the region around $\vec r_0$ the exact densities are well resembled by $n^{\mathrm{Ai}}(\vec r,0)$, which stays positive except very close to ${r=0}$. However, the term in Eq.~(\ref{nAiryTto0}) that is proportional to the Airy function originates in the quantum corrections $\nu_1(x,\vec r)$ and is responsible for the oscillations that render ${n^{\mathrm{Ai}}(\vec r,{T\to0})}$ negative near the potential minimum. As ${r\to0}$, the amplitude and the frequency of these oscillations diverge like ${1/\sqrt{r}}$ and ${1/r}$, respectively. Although the oscillations do not contribute to the particle number, they are clearly an unphysical feature.

Generally, for semiclassical approximations to hold, we require the quantum corrections to be small compared with the classical part. It is obvious from Figs.~\ref{P4finiteT} and \ref{nAiryZeroTAiryZeroTFullRange}, that this condition is neither met for $\nh(\vec r,{T})$ if $T$ is small enough nor for $n^{\mathrm{Ai}}(\vec r,0)$ near positions with ${\nabla U=0}$. To understand the breakdown of the semiclassical gradient expansion in the zero-temperature limit, we now analyze how the unphysical oscillations of $\nh(\vec r,{T})$ and $n^{\mathrm{Ai}}(\vec r,0)$ emerge as $T\to0$.

To take the limit ${T\to0}$ or to choose `small temperatures' for a physical system, we have to identify a characteristic (positive) temperature $T_c$ from the system parameters and consider ${T/T_c\to0}$. Although small temperatures can be defined via the Fermi temperature or excitation energies \textit{before} approximations are made, the semiclassical expansions allows for an ex-post definition of small temperatures: Guided by Fig.~\ref{P4finiteT} and the fact that the TF term of $\nh(\vec r,{T})$ in Eq.~(\ref{nFiniteT}) is always positive, we first take a look at the quantum corrections $\Delta_{\mathrm{qu}}n(\vec r,T)$, whose features as a function of $\vec r$ are dominated by $z/(1+z)^2=\frac14\cosh\big[U/(2k_{\mathrm{B}}T)\big]^{-2}$, and observe that $T$ exclusively appears in the combination $U/(2k_{\mathrm{B}}T)$. Small temperatures in view of $\Delta_{\mathrm{qu}}n(\vec r,T)$ are thus defined by
\begin{align}\label{nSCcriterion}
k_{\mathrm{B}}T\ll\frac{|U(\vec r)|}{2},
\end{align}
with the r.h.s.~vanishing at $\vec r_0$. That is, no temperature $T$ can be considered small in the sense of Eq.~(\ref{nSCcriterion}), at least not uniformly at all $\vec r$. Nonetheless, we can of course choose numerical values of $T$ which obey certain conditions, for example ${k_{\mathrm{B}}T\ll\EF}$.

\begin{figure}
\begin{center}
\includegraphics[width=0.5\linewidth]{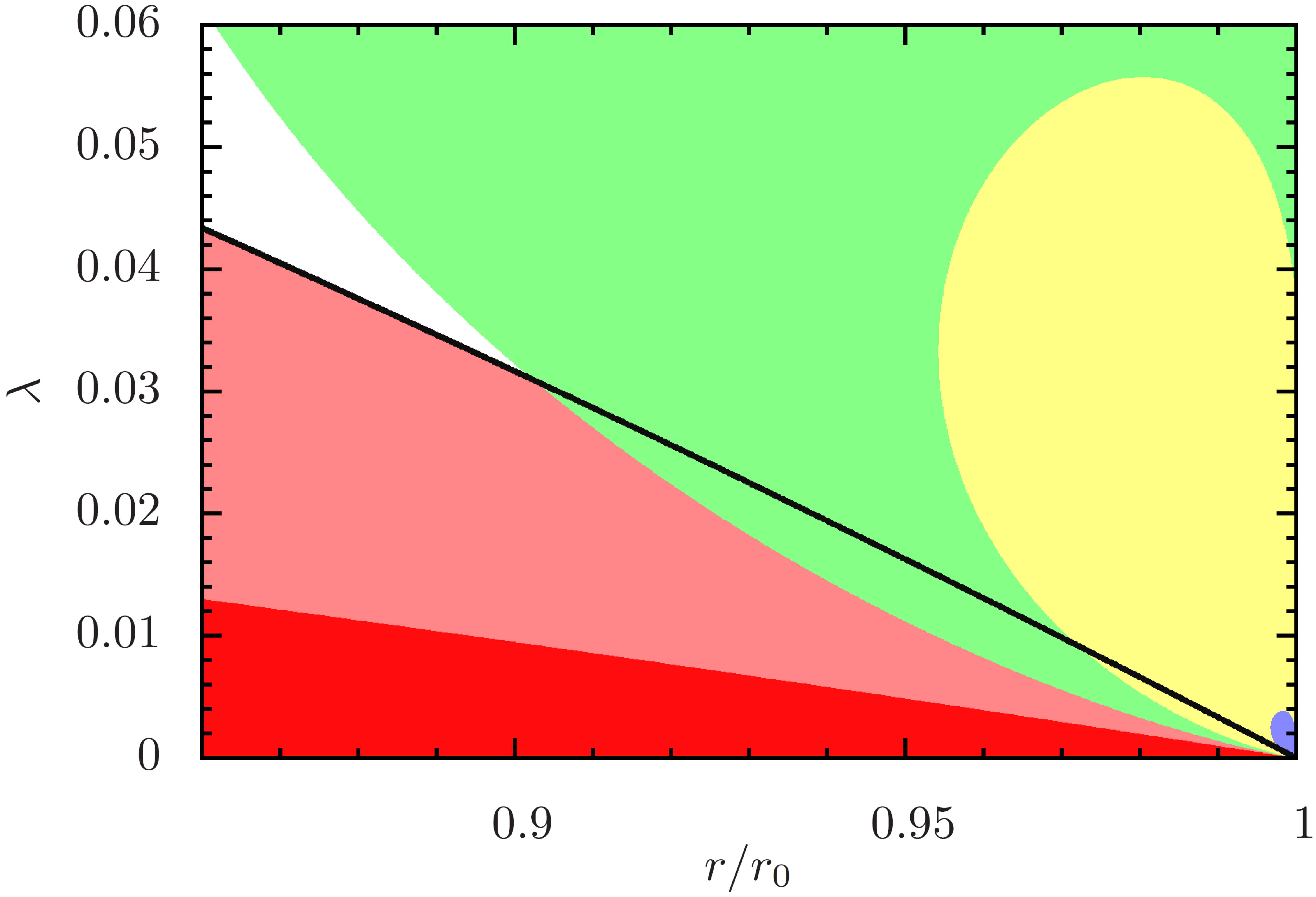}
\caption{\label{NegDensitiesSC}Illustration of the criterion in Eq.~(\ref{nSCcriterion}) heralding the breakdown of the semiclassical approximation $\nh(\vec r,{T})$. The quantum corrections of $n^{\mathrm{sc}}(\vec r,T)$ around $\vec r_0$ are responsible for an unphysical density if ${T=\lambda T_\mathrm{F}}$ is decreased sufficiently. We show areas with ${-\Delta_{\mathrm{qu}}n(\vec r,T)>n^\mathrm{TF}(\vec r,T)}>0$ for ${N=1}$ (green), ${N=42}$ (yellow), and ${N=100172}$ (blue). The light (dark) red area indicates ${k_{\mathrm{B}}T<|U(\vec r)|/6}$ (${k_{\mathrm{B}}T<|U(\vec r)|/20}$), approximately violating (satisfying) Eq.~(\ref{nSCcriterion}). The black line represents ${k_{\mathrm{B}}T=|U(\vec r)|/6}$ and intersects with the regions of negative $n^{\mathrm{sc}}(\vec r,T)$, cf.~Fig.~\ref{P4finiteT}.}
\end{center}
\end{figure}

We declare the range of validity of the semiclassical density by demanding ${\nh(\vec r,{T})>0}$. As illustrated in Fig.~\ref{NegDensitiesSC} for the harmonic oscillator, $\nh(\vec r,{T})$ turns negative in proximity to $\vec r_0$ and for small ${\lambda=T/T_\mathrm{F}}$ if the criterion in Eq.~(\ref{nSCcriterion}) is violated, consistent with the results in Fig.~\ref{P4finiteT}. Given a particle number $N$, we require high enough $T$ and small enough $r$ to guarantee ${n^{\mathrm{sc}}(\vec r,T)>0}$. We observe in Fig.~\ref{NegDensitiesSC} that Eq.~(\ref{nSCcriterion}), e.g., ${k_{\mathrm{B}}T\lesssim|U(\vec r)|/20}$, is a sufficient criterion to define such a region of $\vec r$- and $T$-values.

%..... Of course, the convolution of $K_y(x,{T\to0})$ with $\mathrm{Ai}(x)$ results in a $T$-independent value. But, 
The analysis of the range of validity for $n^{\mathrm{Ai}}(\vec r,T)$ is more involved. Given a small enough $T$, we may consider ${\Delta x\ll 1}$ $\big($see Eq.~(\ref{Deltax})$\big)$ and write
\begin{align}
\int\d x\, \mathrm{Ai}(x)K_y(x,T)\approx\int_{y-\Delta x}^{y+\Delta x}\d x\, \mathrm{Ai}(x)K_y(x,T)\approx\mathrm{Ai}(y)\int_{y-\Delta x}^{y+\Delta x}\d x\, K_y(x,T)=\mathrm{Ai}(y)\frac{\sqrt{2}m^{1/3}}{|\hbar\nab U|^{2/3}},\label{Aiy}
\end{align}
which scales like ${|\nab U(\vec r)|^{-1/2}}$ (since $\mathrm{Ai}(y)$ scales like $|y|^{-1/4}$ for ${-y\gg1}$) and oscillates more and more rapidly due to $\mathrm{Ai}(y)$ for ${\nab U(\vec r)\to0}$, i.e., at potential extrema \footnote{Note that $y<0\Leftrightarrow U<0$, and $y\to-\infty$ corresponds to $|\vec r|\to0$ for the harmonic oscillator.}.
%Although the limit ${T\to0}$ is not performed explicitly in Eq.~(\ref{Aiy}), the $T$-dependence drops out, i.e. the gradient corrections cease to be $T$-dependent for small enough $T$, and
Eq.~(\ref{Aiy}) is reminiscent of the last term in Eq.~(\ref{nAiryTto0}). In contrast to $n^{\mathrm{sc}}(\vec r,T)$, the unphysical features of $n^{\mathrm{Ai}}(\vec r,0)$ do not show up at $\vec r_0$ but at the stationary points of the potential.% While the gradient corrections cease to be $T$-dependent for small enough $T$, the zero-temperature limit is in fact physically ill-defined, as we will argue in the following.

To investigate the zero-temperature limit of $n^{\mathrm{Ai}}(\vec r,T)$, we first observe that the temperature appears in the quantum corrections $\nu_1(x,\vec r)$ only in the combination $(y-x)t_c/T$; see Eq.~(\ref{delta?3}). We thus have to identify ${T_c=|y-x|\,t_c}$ as the characteristic temperature for the case of the Airy-averaged density \footnote{The quantity ${(y-x)}$ is dimensionless, but together with $t_c$ it determines the shape of $K_y(x,T)$ for any given temperature $T$, and, in turn, $K_y(x,T)$ determines the magnitude of the quantum corrections via the integral in Eq.~(\ref{nAirygr0}).}. Hence, the criterion for small $k_{\mathrm{B}}T$, i.e., the analogue of Eq.~(\ref{nSCcriterion}), reads
\begin{align}\label{nAicriterion}
k_{\mathrm{B}}T\ll\frac{|U(\vec r)|}{2}\,|1-x/y|,
\end{align}
equivalent to $\Delta x\ll|x-y|$, and holds if and only if
\begin{align}
K_y(x,T)\ll\underset{x\in\mathbb{R}}{\mathrm{max}}K_y(x,T)=1/(4k_{\mathrm{B}}T).
\end{align}
That is, the $x$-range where Eq.~(\ref{nAicriterion}) is violated coincides with the $x$-range where $K_y(x,T)$ is not negligible. This implies that Eq.~(\ref{nAicriterion}) determines the $x$-range $y\pm\Delta x$ for Eq.~(\ref{Aiy}), which is responsible for negative densities.
%which has to hold \green{uniformly} at all $\vec r$, in particular at $\vec r_0$. Since $x$ takes on arbitrarily small values, no temperature $T$ can be considered small in the sense of Eq.~(\ref{nAicriterion}). This does not mean that the mathematical expression in Eq.~(\ref{delta?3}) is ill-defined at $\vec r_0$ or elsewhere. The limit ${T\to0}$ is mathematically well-defined for all $\vec r$ with ${\nab U(\vec r)\not=0}$, but we loose the means to decide whether a given temperature $T$ is small with respect to the characteristic temperature $T_c$ of the physical system. Nonetheless, we can of course choose numerical values of $T$ which obey certain conditions, for example ${k_{\mathrm{B}}T\ll\EF}$.

\begin{figure}[ht!]
\begin{center}
$\begin{array}{c}
\includegraphics[width=0.5\linewidth]{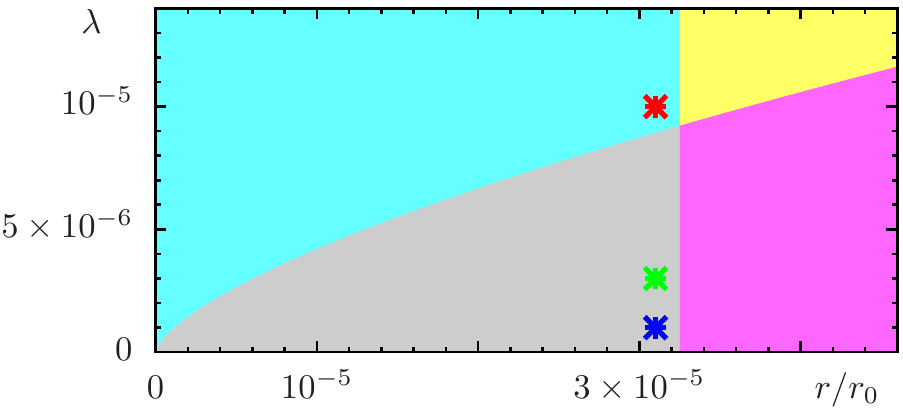}\\
\includegraphics[width=0.5\linewidth]{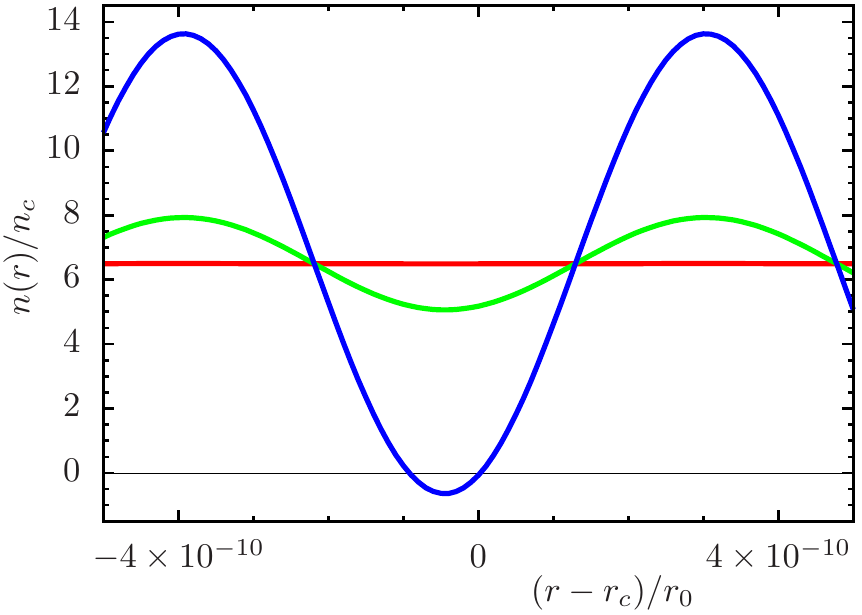}
\end{array}$
\end{center}
\caption{\label{NegDensitiesAi}Upper figure: Negative $n^{\mathrm{Ai}}(\vec r,T)$ can be expected in the gray area that indicates the overlap of the regions in $(r,T)$-space, for which Eqs.~(\ref{Condition2}) [magenta], (\ref{Condition1}) [yellow], and (\ref{Condition1}) [cyan] are satisfied (here for ${N=42}$). We consider Eqs.~(\ref{Condition2}) and (\ref{Condition3}) fulfilled if their l.h.s.~are less than $0.1$ and larger than $10$, respectively. The graph provides an estimate for the regions of validity of $n^{\mathrm{Ai}}(\vec r,T)$, namely, ${T\gtrsim10^{-5}T_\mathrm{F}\approx0.5\,}$pK and ${r\gtrsim3.3\times10^{-5}r_0}$. Lower figure: The Airy-averaged density from Eq.~(\ref{nAirygr0}) for ${N=42}$ particles close to the potential minimum for ${\lambda=10^{-5}}$ (red), ${\lambda=3\times10^{-6}}$ (green), and ${\lambda=10^{-6}}$ (blue). These parameter choices are represented in the upper figure by the colored markings at $r_c=3.1\times10^{-5}r_0$.}
\end{figure}

In view of {Fig.~\ref{nAiryZeroTAiryZeroTFullRange}} and {Eq.~(\ref{Aiy})} we therefore consider small temperatures ${T=\lambda T_\mathrm{F}}$, for ${\lambda\ll 1}$, such that $\Delta x=T\,\mathrm{cosh}^{-1}(\sqrt 2)/t_c$ $\ll 1$, i.e.,
\begin{align}\label{Condition2}
2\,\lambda\, N^{1/3}(R/2)^{-2/3}\ll 1,
\end{align}
where ${0<R=r/r_0<1}$. We also consider positions close to the minimum of the harmonic oscillator potential, i.e., ${-y=|y|\gg1}$ and ${R\ll1}$, which leads to the condition
\begin{align}\label{Condition3}
N^{1/3}|R^{-2/3}-R^{4/3}|\gg 1.
\end{align}
Employing Eq.~(\ref{Aiy}) to compute $n_1^{\mathrm{Ai}}(\vec r,T)$ approximately and replacing $\mathrm{Ai}(y)$ by its approximate amplitude $(\pi^2|y|)^{-1/4}$ for ${-y\gg1}$, we obtain a necessary condition for ${n^{\mathrm{Ai}}(\vec r,T)<0}$, i.e., ${-n_1^{\mathrm{Ai}}(\vec r,T)>n_0^{\mathrm{Ai}}(\vec r,T)>0}$:
\begin{align}\label{Condition1}
36\,\pi\, N^{3/2} \sqrt{R^2-R^4}\lesssim1.
\end{align}
The conditions in Eqs.~(\ref{Condition2})--(\ref{Condition1}) are met simultaneously for specific ranges of $T$ and $r$, depending on the particle number, and illustrated by the gray area in Fig.~\ref{NegDensitiesAi} for ${N=42}$ particles in the harmonic oscillator potential. The overlap of Eqs.~(\ref{Condition2})--(\ref{Condition1}) represents an estimate of an exclusion zone in $(r,T)$-space outside of which $n^{\mathrm{Ai}}(\vec r,T)>0$. The analogous analysis of Eqs.~(\ref{Condition2})--(\ref{Condition1}) for ${N=100172}$ yields the condition $T\gtrsim10^{-15}\,$K that ensures positive $n^{\mathrm{Ai}}(\vec r,T)$ throughout.

%\begin{figure}
%\begin{center}
%$\begin{array}{c}
%\includegraphics[width=0.9\linewidth,height=12em]{ContourPlot_nAi_42}\\
%\includegraphics[width=0.9\linewidth,height=15em]{NegAiryPlot_42_Combined}
%\end{array}$
%\begin{picture}(100,100)
%\put(82.3,263){\textcolor{blue}{\textbf{x}}}
%\put(82.3,277){\textcolor{green}{\textbf{x}}}
%\put(82.3,325){\textcolor{red}{\textbf{x}}}
%\put(120,290){(\ref{Condition2})}
%\put(110,340){(\ref{Condition3})}
%\put(10,320){(\ref{Condition1})}
%\end{picture}
%\end{center}
%\vspace{-12em}
%\caption{\label{NegDensitiesAi}Upper figure \blue{(into Appendix?)}: Negative $n^{\mathrm{Ai}}(\vec r,T)$ can be expected in the black area that indicates the overlap of the regions in $(r,T)$-space, for which Eqs.~(\ref{Condition2})--(\ref{Condition1}) are satisfied (here for ${N=42}$). We consider Eqs.~(\ref{Condition2}) and (\ref{Condition3}) fulfilled if their l.h.s.~are less than $0.1$ and larger than $10$, respectively. The graph provides an estimate for the regions of validity of $n^{\mathrm{Ai}}(\vec r,T)$, namely, ${T\gtrsim10^{-5}T_\mathrm{F}\approx0.5\,}$pK and ${r\gtrsim3.3\times10^{-5}r_0}$. Lower figure: The Airy-averaged density from Eq.~(\ref{nAirygr0}) for ${N=42}$ particles close to the potential minimum for ${\lambda=10^{-5}}$ (red), ${\lambda=3\times10^{-6}}$ (green), and ${\lambda=10^{-6}}$ (blue). These parameter choices are represented in the upper figure by the colored markings. \blue{Produce nicer figures}}
%\end{figure}

For applications that require the zero-temperature particle density, the smooth transition of the Airy-averaged density in Eq.~(\ref{nAiryTto0}) into the classically forbidden region may still be valuable, while the density profiles in the bulk regions near the potential minima are usually well described by the TF density. However, for a semiclassical approximation that is consistent in powers of $\nab$, we have to take into account the gradient corrections for all space. Furthermore, quantities like the ground-state energy obtained from a density functional require the density for all space. That is, the limit ${T\to0}$ is often required for all $\vec r$.

Of course, a priori, unphysical features in the ${T\to0}$ limit of a physical quantity are not necessarily encountered despite an improper physical notion of small temperatures in the sense of violations of Eq.~(\ref{nSCcriterion}) and Eq.~(\ref{nAicriterion}), respectively. However, for the semiclassical particle densities in Eqs.~(\ref{nFiniteT}) and (\ref{nAirygr0}) such unphysical features evidently \textit{do} emerge \footnote{Furthermore, even at $\vec r_0$ the Airy-averaged ground-state density in Eq.~(\ref{nAiryTto0}) is negative if $\frac{\nab^2U(\vec r_0)}{|\nab U(\vec r_0)|^{4/3}}>\left(\frac{81m}{\hbar^2}\right)^{1/3}\frac{(-1/3)!}{(-2/3)!}$, as is easily seen for ${U=0}$. However, for the 2D harmonic oscillator this inequality is only fulfilled for irrelevant values ${N\lesssim0.2}$ of the particle number.}. To conclude, instead of $n^{\mathrm{Ai}}(\vec r,0)$ we have to stick with $n^{\mathrm{Ai}}(\vec r,T)$ which stays well-behaved even for very low but nonzero temperatures, especially across the quantum-classical boundary. Note that this is not an actual restriction for many practical finite-temperature applications that require the finite-temperature densities in the first place.

\section{\label{Kineticenergyfunctional}Semiclassical energy functionals}

In the remainder of this work we investigate the kinetic and total ground-state energies obtained from the semiclassical approximations of the potential functional $E_1[U]$ in Eq.~(\ref{defineE1b}). For reliable approximations of the energy functionals, we demand any approximation of $E_1[U]$ to yield physically well-defined particle densities via Eq.~(\ref{nasdef}). In the previous sections on the semiclassical particle density we found the issues connected to the quantum-classical boundary $\vec r_0$, as outlined in the introduction of this work, to originate in an incompatibility of the zero-temperature limit and the semiclassical gradient expansion. The Airy-averaged functional ${E_1^{\mathrm{Ai}}[U]}$ eliminated the unphysical features of the semiclassical particle density $n^{\mathrm{sc}}(\vec r)$ at $\vec r_0$, derived from ${E_1^{\mathrm{sc}}[U]}$, but led to unphysical oscillations at potential minima --- except for finite temperatures.

Strictly speaking, this means that the total energy and particle density can be consistently obtained in the spirit of DFT from ${E_1^{\mathrm{Ai}}[U]}$ only for finite temperatures, whereas ${E_1^{\mathrm{sc}}[U]}$ also fails for ${T>0}$. However, the numerical values of the semiclassical energies can be checked against the TF and exact energies regardless of the quality of the particle densities. Judging from the performance of the Airy average for the particle density, we expect to gain ground-state energy functionals that can be evaluated unambiguously and that improve on the TF energies.

In Section~\ref{EkinZeroT} we briefly restate the nonzero gradient corrections to the TF kinetic energy in terms of the potential functional $E_1^{\mathrm{sc}}[U]$ as derived in \cite{TrLoNgMuBGE2016}. For the example of the 2D harmonic oscillator we show that approximate ground-state kinetic energies can be derived unambiguously from $E_1^{\mathrm{sc}}[U]$, but not from the density functional $E_{\mathrm{kin}}^{\mathrm{sc}}[n]$.
%As already observed for the particle density in Section~\ref{Inadequacy}, we show that the zero-temperature limit is not a proper physical limit for the Airy-averaged kinetic energy density. However, 
In Section~\ref{AiryPotentialFunctionalNew} we will find the Airy-averaged ground-state functional $E_1^{\mathrm{Ai}}[U]$ to improve on the TF energies unambiguously and consistent with $E_1^{\mathrm{sc}}[U]$.
%From the investigation of $n(\vec r)\cong\nh(\vec r,T)$ in Section~\ref{Oneparticledensity}, the failure of the ${T\to0}$ limit can as well be expected for the semiclassical expansion of the kinetic energy ${E_{\mathrm{kin}}^{\mathrm{sc}}[n,{T\to0}]}$ truncated beyond $\mathcal O(\nab^2)$.

\subsection{\label{EkinZeroT}Ground-state energy functionals}
At ${{T=0}}$ we express the kinetic energy via the functional $E_1[U]$, as defined in Eq.~(\ref{defineE1b}) and approximated according to Eqs.~(\ref{tracef}) and (\ref{fHmu}). For the semiclassical approximation in Eq.~(\ref{fAW}) in 2D, valid up to $\mathcal O(\nab^2)$, we find
%\begin{align}
%E_1[V-\EF ]&\cong E_1^{\mathrm{sc}}[V-\EF]\nn\\
%&=\frac{2}{(2\pi\hbar)^D}\int(\d\vec r)(\d\vec p) \Big[(H_W-\EF )\eta(\EF -H_W)\nn\\
%&\quad+\frac{\hbar^2}{24}\left(\frac{p^2\nab^2V}{D\,m^2}+\frac{(\nab V)^2}{m}\right)\delta'(H_W-\EF )\nn\\
%&\quad+\frac{\hbar^2}{8m}\nab^2V\,\delta(H_W-\EF )\Big]\ ,\label{E1trace}
%\end{align}
\begin{align}
E_1[U]\cong E_1^{\mathrm{sc}}[U]=E_1^{\mathrm{TF}}[U]+\Delta_{\mathrm{qu}} E_1[U] ,\label{E1Ua}
\end{align}
with $E_1^{\mathrm{TF}}[U]=-\frac{m}{2\pi\hbar^2}\int(\d\vec r)\,U^2\,\eta(-U)$ and
\begin{align}\label{deltaquE1}
\Delta_{\mathrm{qu}} E_1[U]=\frac{1}{24\pi}\int(\d\vec r)\,(\nab^2U)\,\eta(-U) .
\end{align}
That is, there \textit{are} nonzero and unambiguous quantum corrections beyond the TF approximation in 2D. For $U(\vec r)=\frac12 m\omega^2\vec r^2-\mu$ they amount to ${\Delta_{\mathrm{qu}} E_1[U]=\mu/6}$.
 
Following \cite{TrLoNgMuBGE2016}, using Eq.~(\ref{nasdef}), and consistently neglecting derivatives of the potential beyond second order, we obtain the semiclassical ground-state density
\begin{align}\label{nDeltan}
\nh(\vec r)&=n^{\mathrm{TF}}(\vec r)+\Delta_{\mathrm{qu}} n(\vec r)
\end{align}
from the total variation $\delta E_1^{\mathrm{sc}}[U]$ of Eq.~(\ref{E1Ua}), with $n^{\mathrm{TF}}(\vec r)$ from Eq.~(\ref{TFdensity}), and the leading gradient corrections
\begin{align}\label{deltaqn}
\Delta_{\mathrm{qu}} n(\vec r)=\frac{1}{24\pi}\Big(\nab^2\eta(-U)-(\nab^2U)\,\delta(U)\Big),
\end{align}
which are nonzero only at the quantum-classical boundary $\vec r_0$, consistent with the low-temperature behavior of $\nh(\vec r,T)$ discussed in Section~\ref{Semiclassicalparticledensity}.

Equations (\ref{E1Ua}) and (\ref{nDeltan}) establish the semiclassical expansions of ${E_1[U]}$ and $n(\vec r)$ up to $\mathcal O(\nab^2)$. To express the kinetic energy $E_{\mathrm{kin}}$ as an approximate density functional $E_{\mathrm{kin}}^{\mathrm{sc}}[n]$, the $U$-dependence of $n$ in Eq.~(\ref{nDeltan}) has to be inverted term by term, and the resulting $n$-dependent effective potential energy is inserted into Eq.~(\ref{E1Ua}). This perturbative evaluation reproduces the gradient corrections of $E_{\mathrm{kin}}^{\mathrm{sc}}[n]$ in 1D and 3D, reported in  Table~\ref{IntroTable}. As argued in \cite{TrLoNgMuBGE2016}, this elimination of the potential in favor of the particle density leads to an ambiguous kinetic-energy density functional $E_{\mathrm{kin}}^{\mathrm{sc}}[n]$ at $\mathcal O(\nab^2)$: The explicit $n$-dependence of the leading term $U_0$ of $U\cong U_0+\Delta_{\mathrm{qu}} U$, where $\Delta_{\mathrm{qu}} U$ is the $\mathcal O(\nab^2)$ correction to $U_0$, cannot be obtained from Eq.~(\ref{nDeltan}) for all space.
%This is also the case for the quantum corrections $\Delta_{\mathrm{qu}} U(\vec r)=-\frac{\pi\hbar^2}{m}\Delta_{\mathrm{qu}} n(\vec r)$ of $U(\vec r)$, which are obtained from Eq.~(\ref{nDeltan}) by inserting $U_0$ into $\Delta_{\mathrm{qu}} n(\vec r)$.
Inside the classically allowed region, where ${U<0}$ and $\Delta_{\mathrm{qu}} n(\vec r)=0$, Eq.~(\ref{nDeltan}) is easily inverted,
\begin{align}
U=U_0=-\frac{\pi\hbar^2}{m}n.
\end{align}
The classically forbidden region is defined as ${U>0}$, and this information about $U$ is all we can get from $\nh(\vec r)$ due to the step function in $n^{\mathrm{TF}}(\vec r)$. For ${U=0}$ the gradient correction in Eq.~(\ref{deltaqn}) is singular, and the semiclassical $\mathcal O(\nab^2)$ approximation is not permissible in the first place.

The inversion of the $U$-dependence of $n$ is possible in the classically allowed region, that is, at the TF level. However, the inversion cannot be obtained for all space and, hence, does not allow for a density functional $E_{\mathrm{kin}}^{\mathrm{sc}}[n]$. If we disregard this observation and insert the leading term $U_0$ into Eq.~(\ref{E1Ua}), we find
\begin{align}
E_{\mathrm{kin}}^{\mathrm{sc}}[n]&=E_1^{\mathrm{TF}}[U]+\Delta_{\mathrm{qu}} E_1[U]-\int(\d\vec r)\,U(\vec r)\, n(\vec r)=E_{\mathrm{kin}}^{\mathrm{TF}}[n]+\Delta_{\mathrm{qu}} E_{\mathrm{kin}}[n] ,\label{newEkin2D}
\end{align}
where ${E_{\mathrm{kin}}^{\mathrm{TF}}[n]=\int(\d\vec r)\frac{\pi\hbar^2}{2m}n^2}$ is the TF kinetic-energy density functional in 2D, and
\begin{align}
\Delta_{\mathrm{qu}} E_{\mathrm{kin}}[n]=\int(\d\vec r)\frac{\hbar^2}{24m}\delta(n)(\nab n)^2\label{gradcorr2D}
\end{align}
may be regarded as a candidate for the leading-order quantum corrections beyond the TF approximation. Here, we consistently keep only terms up to $\mathcal O(\nab^2)$.% In fact, the quantum corrections of $E_1^{\mathrm{TF}}[U]$ coming from $\Delta_{\mathrm{qu}} U$ exactly cancel those of $\int(\d\vec r)\,U\big(n(\vec r)\big)\, n(\vec r)$.

The gradient corrections in Eq.~(\ref{gradcorr2D}) coincide with those given in \cite{DissvanZyl,Brack2003}. There, $\Delta_{\mathrm{qu}} E_{\mathrm{kin}}[n]$ is reported to vanish since the exact density is argued to be nonzero everywhere, such that $\delta(n)$ yields ${\Delta_{\mathrm{qu}} E_1[U]=0}$, which is a plausible argument. But, if Eq.~(\ref{newEkin2D}) were a consistent approximation up to $\mathcal O(\nab^2)$, we could insert the TF density into Eq.~(\ref{newEkin2D}) to obtain a kinetic energy consistent up to $\mathcal O(\nab^2)$. For such a perturbation-theoretic procedure, $n^{\mathrm{TF}}(\vec r)$ could yield nonzero gradient corrections, since it vanishes at $\vec r_0$, while its gradient in the classically allowed region is nonzero. However, we demonstrate in the following for the example of the 2D harmonic oscillator that such an evaluation of Eq.~(\ref{newEkin2D}) is ambiguous. In other words, $E_{\mathrm{kin}}^{\mathrm{sc}}[n]$ in Eq.~(\ref{newEkin2D}) is not well-defined in the first place.

%\blue{In other words, a kinetic energy which is consistent up to $\mathcal O(\nab^2)$ may also be obtained from Eq.~(\ref{newEkin2D}) by inserting the semiclassical density up to $\mathcal O(\nab^2)$ into $E_{\mathrm{kin}}^{\mathrm{TF}}[n]$ and the TF density, which is of order $\nab^0$, into $\Delta_{\mathrm{qu}} E_{\mathrm{kin}}[n]$}\footnote{\blue{We may also insert the TF density into both the leading term and the corrections since first-order variations of the energy have to vanish anyway?}}.

%\blue{Since the TF density \textit{is} zero at the classical turning points, $\Delta_{\mathrm{qu}} E_1[U]$ could in fact be nonzero. We elaborate on this possibility for the example of a 2D harmonic oscillator in the next section.}

%\subsection{\label{Ekin2DHO}Harmonic oscillator:\\ Ambiguous density functional in 2D}
%\blue{Shorter, more words, less equations.}
%In this section we assemble inconsistencies for the example of the 2D harmonic oscillator, which arise if the kinetic energy is calculated from Eq.~(\ref{newEkin2D}), thereby underpinning that Eq.~(\ref{newEkin2D}) is not well-defined in the first place since the elimination of the potential in favor of the particle density is restricted to the classically allowed region.

\begin{table}
\begin{center}
\begin{tabular}{c|c|c|c|c} \toprule
 & $\Delta_{\mathrm{qu}} E_1(N)$; Eq.~(\ref{deltaquE1}) & \multicolumn{3}{c}{$2\,\Delta_{\mathrm{qu}} E_{\mathrm{kin}}$; Eq.~(\ref{gradcorr2D})}\\
 &  & ${\nab n^{\mathrm{TF}}=0}$ & \multicolumn{2}{c}{${\nab n^{\mathrm{TF}}\not=0}$}\\ \midrule
$c$ & $1/6$ & $0$ & $\eta(0)/3$ & $1/6$\\ \bottomrule
\end{tabular}
\end{center}
\caption{\label{GradCorrInconsistencies}Leading gradient corrections (in units of $\hbar\omega\sqrt{N}$) to the TF energy from the perturbative evaluations of Eqs.~(\ref{E1Ua}) and (\ref{newEkin2D}). The leading gradient correction of the exact energy in Eq.~(\ref{ENexact}) amounts to  ${\bar{c}=1/4}$ on average. The cusp of the TF density at the quantum-classical boundary prevents an unambiguous evaluation of Eq.~(\ref{newEkin2D}). Although there is no unambiguous way of assigning a value to $\eta(0)$, we can avoid the explicit evaluation of the step function via an integration by parts of $\Delta_{\mathrm{qu}} E_{\mathrm{kin}}[n^{\mathrm{TF}}]$ and find $c=1/6$. This finding is consistent with $c=\eta(0)/3$ only for ${\eta(0)=1/2}$, whereas the average exact coefficient $\bar{c}$ is recovered only for the uncommon choice ${\eta(0)=3/4}$.}
\end{table}

In Table~\ref{GradCorrInconsistencies} we give the particle-number scalings of the $\mathcal O(\nab^2)$ approximations obtained from Eqs.~(\ref{E1Ua}) and (\ref{newEkin2D}), respectively. The exact energy of the 2D harmonic oscillator for continuous $N$ is
\begin{align}\label{ENexact}
\frac{E(N)}{\hbar\omega}=\frac{E^{\mathrm{TF}}(N)}{\hbar\omega}+c\,N^{1/2}+\mathcal O(N^{-1/2}) ,
\end{align}
with ${E^{\mathrm{TF}}(N)=2\,E_{\mathrm{kin}}^{\mathrm{TF}}[n^{\mathrm{TF}}]=\frac23\hbar\omega N^{3/2}}$ and $c$ an oscillatory function of $N$, cf.~endnote [63] in \cite{TrLoNgMuBGE2016}. The semiclassical approximations give an average account of the exact shell oscillations, cf.~Fig.~1 in \cite{TrLoNgMuBGE2016}.

%The coefficient ${c(s)=1/3-s^2}$ involves the function $s=s(\sigma,N)=\sqrt{N+1/4}-\sigma-1$ with the number $\sigma$ of filled shells, which interpolates between filled shells of the 2D harmonic oscillator and yields ${\bar{c}=1/4}$ on average \footnote{We have ${s(\sigma+1,N(\sigma+1,\kappa))=s(\sigma,N(\sigma,\kappa))}$, where $N$ depends on $\sigma$ and on the fractional filling ${0\le\kappa<1}$ of the highest occupied shell. Furthermore, ${-1/2\le s<1/2}$, where ${s=-1/2}$ corresponds to completely filled shells, and ${\int_{-1/2}^{1/2}\d s\,s^2=1/12}$, such that the average of $c(s)$ is ${\bar{c}=1/4}$.}. The semiclassical approximations give an average account of the exact shell oscillations, cf.~Fig.~1 in \cite{TrLoNgMuBGE2016}.

When evaluating Eq.~(\ref{E1Ua}) perturbatively \footnote{With the ground-state variables in TF approximation, $V^{\mathrm{TF}}=\frac12m\omega^2\vec r^2$ and $\mu^{\mathrm{TF}}=\hbar \omega\sqrt{N}$, we have $E^{\mathrm{sc}}(N)=E^{\mathrm{sc}}[V^{\mathrm{TF}},\mu^{\mathrm{TF}}]=E_1^{\mathrm{TF}}[V^{\mathrm{TF}}-\mu^{\mathrm{TF}}]+\Delta_{\mathrm{qu}}E_1[V^{\mathrm{TF}}-\mu^{\mathrm{TF}}]+\mu^{\mathrm{TF}}N$. Observe that ${\mu^{\mathrm{sc}}=\hbar\omega\sqrt{N+1/6}}$ is the appropriate chemical potential that yields the particle number $N$ from the spatial integral of $\nh(\vec r)$ in Eq.~(\ref{nDeltan}), consistent with the fact that $E^{\mathrm{sc}}$ as a function of $\mu$ coincides with the TF expression ${E^{\mathrm{TF}}(\mu)=\frac23\frac{\mu^3}{(\hbar\omega)^2}}$, i.e., $E^{\mathrm{sc}}(\mu^{\mathrm{sc}})$ coincides with Eq.~(\ref{Ehbar2}) up to $\mathcal O(N^{1/2})$.
}, we obtain the total energy
\begin{align}\label{Ehbar2}
E^{\mathrm{sc}}(N)=E^{\mathrm{TF}}(N)+\Delta_{\mathrm{qu}} E_1(N),
\end{align}
with $\Delta_{\mathrm{qu}} E_1(N)=(\hbar\omega/6)\,N^{1/2}$; see Table~\ref{GradCorrInconsistencies}. Since $E^{\mathrm{sc}}(N)$ obeys the virial theorem, the semiclassical potential functional in Eq.~(\ref{E1Ua}) unambiguously yields a \textit{nonzero} gradient correction ${\hbar\omega\sqrt{N}/12}$ to $E_{\mathrm{kin}}$.

%In contrast, we make the following observations for the gradient correction $\Delta_{\mathrm{qu}} E_{\mathrm{kin}}[n^{\mathrm{TF}}]$ of the density functional in Eq.~(\ref{newEkin2D}). Employing
%\begin{align}\label{nabnTF}
%\big(\nab n^{\mathrm{TF}}(\vec r)\big)^2=\left(\frac{m\omega}{\hbar\sqrt{\pi}}\right)^4r^2\,\eta(r_0^2-r^2) ,
%\end{align}

%\left.\frac{E^{\mathrm{sc}}(N)}{\hbar\omega}\right|_{\mathrm{from Eq.}\;(\ref{newEkin2D})}

In contrast, we get no quantum corrections from the kinetic-energy \textit{density} functional in Eq.~(\ref{newEkin2D}) up to $\mathcal O(\nab^2)$ if we employ ${\left.\nab n^{\mathrm{TF}}(\vec r)\right|_{\vec r=\vec r_0}=0}$, whereas $2\,\Delta_{\mathrm{qu}} E_{\mathrm{kin}}[n^{\mathrm{TF}}]\not=0$ for ${\left.\nab n^{\mathrm{TF}}(\vec r)\right|_{\vec r=\vec r_0}\not=0}$ \footnote{The virial theorem employed at the level of the TF approximation also holds for the energy to $\mathcal O(\nab^2)$ that includes the approximate kinetic energy in Eq.~(\ref{newEkin2D}), as can be checked via a scaling transformation along the lines presented in \cite{FangBerge2011}.}. %\footnote{Observe that the step function in Eq.~(\ref{nabnTF}) is a viable test function for the $\delta$ function since it has compact support in the variables $r$ and $r^2$, respectively.}.

%Using $\delta(\lambda)\eta(\lambda)=\frac12\partial_\lambda\eta^2(\lambda)$ in Eq.~(\ref{DquEkin1}), we can avoid the explicit evaluation of the step function via an integration by parts and find
%\begin{align}\label{DquEkin2}
%\frac{2}{\hbar\omega}\Delta_{\mathrm{qu}} E_{\mathrm{kin}}[n^{\mathrm{TF}}]=\frac{1}{6} N^{1/2}.
%\end{align}

%We now compare the energy corrections in Eqs.~(\ref{Ehbar2}), (\ref{DquEkin1}), and (\ref{DquEkin2}) to the leading corrections of 

%Equation.~(\ref{DquEkin1}) recovers the average coefficient $\bar{c}$ only for the uncommon choice ${\eta(0)=3/4}$, but is consistent with Eq.~(\ref{DquEkin2}) only for ${\eta(0)=1/2}$. %With ${\nab n^{\mathrm{TF}}(\vec r)=0}$ in the classically forbidden region, we get vanishing quantum corrections .
Evidently, Eq.~(\ref{newEkin2D}) does not incorporate enough information from the classically forbidden region to retrieve the kinetic-energy beyond the TF approximation consistently. Contrary to this, the potential functional $E_1^{\mathrm{sc}}[U]$ in Eq.~(\ref{E1Ua}) is well-defined for all space and provides unambiguous quantum corrections. However, the corresponding singular particle density in Eq.~(\ref{nDeltan}) remains unsatisfactory. In the remainder of this work we therefore revert to the Airy-averaging method, which provides a smooth transition of the particle density into the classically forbidden region, and investigate its quality regarding the semiclassical energies.

\subsection{\label{AiryPotentialFunctionalNew}Airy-averaged density-potential functional}

In Section~\ref{EkinZeroT} we used the Wigner transform in Eq.~(\ref{fAW}) to calculate the potential functional ${E_1^{\mathrm{sc}}[U]}$; see Eq.~(\ref{E1Ua}). The elimination of the potential in favor of the particle density led to a troublesome kinetic-energy density functional $E_{\mathrm{kin}}^{\mathrm{sc}}[n]$ which suffered from ambiguities related to the quantum-classical boundary $\vec r_0$ --- in contrast to the potential functional $E_1^{\mathrm{sc}}[U]$ that incorporates information from the classically forbidden region. Motivated by the high quality of the Airy-averaged particle density in Eq.~(\ref{nAiryTto0}) in the vicinity of $\vec r_0$, we now compare the Airy average of the ground-state potential functional $E_1[U]$ with $E_1^{\mathrm{sc}}[U]$, the exact energy, and the TF energy.

From the Airy-averaged expression in Eq.~(\ref{trfA4}) we get
\begin{align}
E_1[U]\cong E_1^{\mathrm{Ai}}[U]&=\int(\d\vec r)\left[\frac{\nab^2U}{12\pi}\mathcal A(y)-\frac{m^{1/3}|\hbar\nab U|^{4/3}}{8\pi\hbar^2}\big(y^2\mathcal A(y)+\mathrm{Ai}(y)+y\mathrm{Ai}'(y)\big)\right] .\label{E1AiryNew}
\end{align}
As a consistency check the particle density in Eq.~(\ref{nAiryTto0}) can be derived from Eq.~(\ref{E1AiryNew}) via Eq.~(\ref{nasdef}). The variation of $E_1^{\mathrm{Ai}}[U]$ \wrt $U$ includes terms which stem from the variation of $\nab U$ \wrt $U$. However, these terms come with derivatives of $U$ beyond second order that can be neglected at the level of our $\mathcal O(\nab^2)$ approximation. That is, we may obtain $\delta E_1^{\mathrm{Ai}}[U]$ at $\mathcal O(\nab^2)$ by varying Eq.~(\ref{E1AiryNew}) only \wrt $U$, not \wrt $\nab U$. Equivalently, for arbitrary operator-valued functions $f(H)$ we may use the relation ${\delta\hspace{0.1ex}\tr\{f(H)\}=\tr\{f'(H)\delta H\}}$, where only terms linear in ${[\delta H]_W(\vec r,\vec p)=\delta U(\vec r)}$ are taken into account. Then, the terms from varying $\nab U$ \wrt $U$ do not show up in the first place. In either case we obtain the Airy-averaged particle density in Eq.~(\ref{nAiryTto0}) via Eq.~(\ref{nasdef}). Note that it does not seem to be feasible to express $U$ as a function of the particle density by inverting Eq.~(\ref{nAiryTto0}), which would be required for obtaining a kinetic-energy density-functional $E_{\mathrm{kin}}[n]$ from $E_1[V-\mu]$, cf.~Eq.~(\ref{defineE1b}).

We now assess the quality of the quantum corrections in Eq.~(\ref{E1AiryNew}) as a function of $N$ for the noninteracting 2D harmonic oscillator. At the level of the approximation in Eq.~(\ref{E1AiryNew}) the chemical potential $\mu^{\mathrm{Ai}}$ is calculated from Eq.~(\ref{getmuAi}) for a given particle number $N$. We find ${\mu^{\mathrm{Ai}}=\mu^{\mathrm{sc}}}$ with sufficient numerical accuracy. With Eqs.~(\ref{EnergyVnmu}) and (\ref{E1AiryNew}) the total Airy-averaged energy for noninteracting systems is expressed as
\begin{align}
E^{\mathrm{Ai}}&=E_1^{\mathrm{Ai}}+\EF^\mathrm{Ai} N=-\int(\d\vec r)\left\{\tau^{\mathrm{Ai}}(\vec r)-\frac{\nab^2U}{24\pi}\mathcal A(y)\right\}+\EF^\mathrm{Ai} N\label{EtotAiNew}\\
&=2\int(\d\vec r)\,\tau^{\mathrm{Ai}}(\vec r) ,\label{EtotAi2New}
\end{align}
where the kinetic energy density can be written as
\begin{align}
\tau^{\mathrm{Ai}}(\vec r)&=\frac{m^{1/3}|\hbar\nab U|^{4/3}}{8\pi\hbar^2}\Big[y^2\mathcal A(y)+\mathrm{Ai}(y)+y\mathrm{Ai}'(y)\Big]-\frac{\nab^2U}{24\pi}\mathcal A(y). \label{tauT0nabnot0}
\end{align}
While Eq.~(\ref{EtotAiNew}) holds in general, we used the virial theorem for Eq.~(\ref{EtotAi2New}), leading to the numerically more feasible expression
\begin{align}\label{EtotAi2New2}
E^{\mathrm{Ai}}=\frac23\EF^\mathrm{Ai} N+\int(\d\vec r)\frac{\nab^2U}{36\pi}\mathcal A(y) .
\end{align}

Table \ref{TableEnergiesNew} illustrates our results for the semiclassical total energies for various particle numbers $N$. We compare the exact energies $E^{\mathrm{ex}}$ with the TF energy $E^{\mathrm{TF}}$, the semiclassical energies $E^{\mathrm{sc}}$ and the Airy-averaged energies $E^{\mathrm{Ai}}$. The virial theorem and an integration by parts lead to the numerically more feasible expression
\begin{align}\label{Ehbar2T0}
E^{\mathrm{sc}}=2E_{\mathrm{kin}}^{\mathrm{sc}}=\int(\d\vec r)\,\frac{m}{\pi\hbar^2}U^2\eta(-U),
\end{align}
used as a consistency check of Eq.~(\ref{Ehbar2}). The leading gradient corrections are implicitly included in Eq.~(\ref{Ehbar2T0}) via the semiclassical particle density to $\mathcal O(\nab^2)$ that determines the chemical potential ${\mu=\mu^{\mathrm{sc}}\not=\mu^{\mathrm{TF}}}$. Both expressions of $E^{\mathrm{Ai}}$, Eqs.~(\ref{EtotAiNew}) and (\ref{EtotAi2New2}), converge to $E^{\mathrm{sc}}$ with increasing numerical accuracy for increasing $N$. Note that the semiclassical energies $E^{\mathrm{TF}}$, $E^{\mathrm{sc}}$, and $E^{\mathrm{Ai}}$ all have errors of about the same size when compared with the exact energies. The Airy-averaged quantities improve upon the TF approximation mostly in the particle density, not so much in the energy.

\begin{table}
\begin{center}
\begin{tabular}{cccc} \toprule
$N$ & $E^{\mathrm{ex}}\; [\hbar\omega]$ & $\epsilon^{\mathrm{TF}}$ & $\epsilon^{\mathrm{Ai}}$; $\epsilon^{\mathrm{sc}}$ \\ \midrule
42 & 182 & $-1.5\times10^{-3}$ & $1.5\times10^{-3}$ \\
168.5 & 1462.5 & $-1.5\times10^{-3}$ & $-0.7\times10^{-3}$ \\
1054 & 22816 & $-8\times10^{-5}$ & $4\times10^{-5}$ \\
4216.5 & 182552.5 & $-6\times10^{-5}$ & $-3\times10^{-5}$ \\
100172.3 & 21136387 & $-6\times10^{-7}$ & $6\times10^{-7}$ \\ \bottomrule
\end{tabular}
\end{center}
\caption{\label{TableEnergiesNew}Approximate values of the normalized differences ${\epsilon^{\mathrm{Ai}}=(E^{\mathrm{Ai}}-E^{\mathrm{ex}})/(E^{\mathrm{Ai}}+E^{\mathrm{ex}})}$ (and, analogously, $\epsilon^{\mathrm{sc}}$ and $\epsilon^{\mathrm{TF}}$) for various particle numbers, with the TF energy $E^{\mathrm{TF}}$, $E^{\mathrm{sc}}$ from Eq.~(\ref{Ehbar2}), and $E^{\mathrm{Ai}}$ from Eq.~(\ref{EtotAi2New2}). The semiclassical energy functionals allow for non-integer $N$, in which case $E^{\mathrm{ex}}$ is taken as a linearly weighted sum of exact energies. We find $\epsilon^{\mathrm{Ai}}\approx\epsilon^{\mathrm{sc}}$ with high numerical accuracy.}
\end{table}

%\begin{table*}
%\centering
%\caption{\label{TableEnergiesNew}Comparison of the exact energies $E^{\mathrm{ex}}$ for several particle numbers $N$ of the 2D harmonic oscillator with the TF energy $E^{\mathrm{TF}}$, the semiclassical energies $E^{\mathrm{sc}}$, and the Airy-averaged energies $E^{\mathrm{Ai}}$. All energies are in units of $\hbar\omega$. \blue{I included noninteger particle numbers to emphasize explicitly that we are not restricted to integer particle numbers.}}
%\begin{ruledtabular}
%\setlength\extrarowheight{0.5em}
%\begin{tabular}{cccccc}
%$N$ & 42 & 168.5 & 1054 & 4216.5 & 100172.3 \\
%\hline\\[-1.4em]
%$E^{\mathrm{ex}}$ & 182 & 1462.5 & 22816 & 182552.5 & 21136387 \\
%$E^{\mathrm{TF}}$ & 181.4607 & 1458.1715 & 22812.331 & 182531.11 & 21136361 \\
%$E^{\mathrm{sc}}$; Eq.~(\ref{Ehbar2}) & 182.5409 & 1460.3349 & 22817.741 & 182541.93 & 21136413 \\
%$E^{\mathrm{sc}}$; Eq.~(\ref{Ehbar2T0}) & 182.5419 & 1460.3355 & 22817.741 & 182541.93 & 21136413 \\
%$E^{\mathrm{Ai}}$; Eq.~(\ref{EtotAiNew}) & 182.6137 & 1460.3435 & 22817.741 & 182542.14 & 21136415 \\
%$E^{\mathrm{Ai}}$; Eq.~(\ref{EtotAi2New2}) & 182.5414 & 1460.3354 & 22817.742 & 182541.94 & 21136413
%\end{tabular}
%\end{ruledtabular}
%\end{table*}

\begin{figure}[htb!]
\begin{center}
\includegraphics[width=0.5\linewidth]{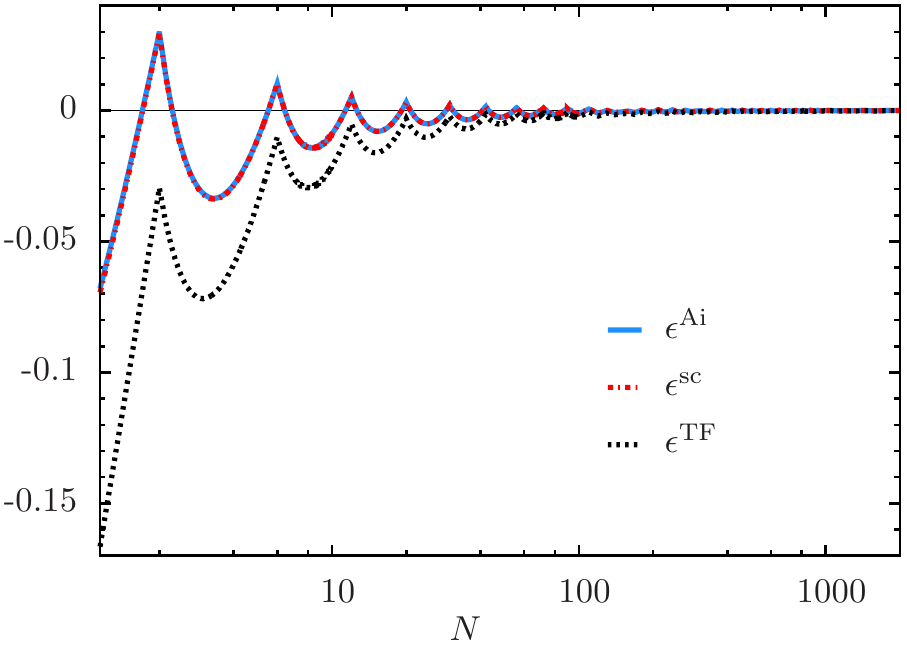}
\caption{\label{EkinAiry20160420EnergyComparisonLogPlot}Comparison of $E^{\mathrm{sc}}$ from Eq.~(\ref{Ehbar2}) and $E^{\mathrm{Ai}}$ from Eq.~(\ref{EtotAi2New2}) with the TF energy $E^{\mathrm{TF}}$. We show the normalized differences $\epsilon$ as defined in Table \ref{TableEnergiesNew} for particle numbers $N$ up to 2000. The trend of decreasing amplitude of oscillations continuous for larger $N$. For instance, particle numbers around $10^6$ yield $\epsilon^{\mathrm{Ai}}\sim\mathcal O(10^{-7})$.}
\end{center}
\end{figure}

In Fig.~\ref{EkinAiry20160420EnergyComparisonLogPlot} we compare our semiclassical approximations with the exact and TF energies. While the TF approximation performs as well as the semiclassical energies for completely filled oscillator shells, we generally find $E^{\mathrm{Ai}}$ and $E^{\mathrm{sc}}$ to outperform $E^{\mathrm{TF}}$.

\FloatBarrier

\section{Conclusions and perspectives}
In the present work we addressed semiclassical kinetic-energy functionals and particle densities, in particular targeting systems with reduced dimensions. As reported thus far, the leading $\mathcal O(\nab^2)$ gradient corrections beyond the Thomas-Fermi (TF) approximation of the density-only functionals in 1D and 2D proved to be ambiguous in 2D and not even bounded from below in 1D. In 2D we showed explicitly that the incomplete inversion of the TF particle density as a function of the potential lies at the heart of the trouble with the $\mathcal O(\nab^2)$ gradient corrections. We showed how information from the classically forbidden region is conveniently incorporated by keeping the (effective) potential, which generally includes a contribution from the interactions, as a variable of the total energy functional. Thereby, we were able to derive kinetic-energy functionals beyond the TF approximation that yield unambiguous energies in 2D.

These improvements upon the TF approximation were achieved by rewriting the total energy functional in terms of a potential-only functional $E_1$ that comes with two advantageous properties. Its expression as a single-particle trace of an operator function of the effective one-particle Hamiltonian allowed for a systematic semiclassical expansion, viz.~semiclassical approximation, with the aid of Wigner transforms, and its functional derivative directly yields the particle density. We demonstrated that the well-known Wigner function to order $\nab^2$ coincides with the two leading terms of a Wigner function that is approximated by an average over Airy functions. Thereby, higher orders of $\nab$ are included, and a smooth transition of observables into the classically forbidden regime is obtained. This smooth transition proves necessary for the particle density since its approximation derived from $E_1$ at $\mathcal O(\nab^2)$ is physically ill-defined as it becomes negative near the quantum-classical boundary, even for finite temperatures. In contrast, the particle density obtained from the Airy-averaged $E_1$ stays positive near the quantum-classical boundary and agrees excellently with the exact densities for both finite and zero temperature --- we presented data for the 2D harmonic oscillator. However, the Airy-averaged particle density exhibits unphysical oscillations in the vicinity of stationary points of the effective potential, where the TF approximation is supposed to work particularly well. Though, in contrast to the particle density at $\mathcal O(\nab^2)$, the Airy-averaged particle density at very low but finite temperatures provides an excellent approximation to the exact density for all space and should prove valuable for experiments with ultracold Fermi gases that require the density distribution across the quantum-classical boundary.

We traced back the unphysical behavior of the semiclassical particle densities to a problematic zero-tem\-per\-a\-ture limit. The semiclassical expressions for the finite-temperature particle densities involve exactly one characteristic temperature that can be viewed as a thermometer. This characteristic temperature depends on the potential in such a way that small temperatures cannot be defined uniformly at all spatial positions. We thus demonstrated that the zero-temperature limit, although mathematically well-defined, is physically incompatible with the employed semiclassical approximations of the particle density. On the other hand, the Airy-averaged semiclassical energies exhibit no unphysical characteristics. Expanding the discussion in \cite{TrLoNgMuBGE2016}, we showed that the energies obtained from the density-potential functional approach include nonzero gradient corrections and improve upon the TF approximation.%
%in both 1D and 2D. 
%We thus resolved the long-standing dilemma of vanishing gradient corrections to the two-dimensional kinetic energy functional and opened the door to much more accurate descriptions of one- and two-dimensional interacting many-body systems than hitherto available.
%Our results for the kinetic energy were corroborated by a discussion of the kinetic energy density that served as an additional check of the methods developed in this work.

We intend to apply our methods to specific interacting systems like a two-component Fermi gas with contact and dipole-dipole interactions, thereby going beyond the treatment in \cite{KazikBGE2001,FangBerge2011}. Furthermore, harmonic-oscillator eigenfunctions, which could be introduced along the same lines as the Airy average, may prove useful in obtaining semiclassical approximations with a well-defined zero-temperature limit while retaining the improvement over the TF approximation.

\section*{Acknowledgements}
We sincerely thank Cord~Axel~M\"uller for his continuous support, countless valuable discussions, and insights. His feedback helped much in clarifying many details. This work is funded by the Singapore Ministry of Education and the National Research Foundation of Singapore. H.K.N.~is also funded by a Yale-NUS College start-up grant.
%C.A.M.~acknowledges the hospitality of the Institut Non Lin\'eaire de Nice (CNRS and Universit\'e de Nice) and the Laboratoire de Physique Th\'eorique (CNRS and UPS Toulouse).

%\section*{\red{Appendix}}

\setcounter{section}{0}

\renewcommand{\theequation}{\thesection.\arabic{equation}}
\setcounter{equation}{0}

%\section{\label{WignerAppendix}Wigner's semiclassical expansion}
%In this appendix we collect the basic quantities of Wigner's phase-space formulation insofar as they are needed for the present work and give the derivations of the semiclassical expressions in Eqs.~(\ref{fAW}) and (\ref{trfA4}). %\blue{Reformulate Appendices \ref{WignerphasespaceAppendix} and \ref{SemiclassicalexpansionofWignertransforms}? (for now they are more or less copied from the draft of the spectral-function-paper with Cord)}

\renewcommand{\thesection}{Appendix \Alph{section}}
\section{\label{WignerphasespaceAppendix}Wigner's phase-space formulation}
\renewcommand{\thesection}{\Alph{section}}

In this appendix we collect the basic quantities of Wigner's phase-space formulation insofar as they are needed for the present work. For $D$ spatial dimensions the trace of any operator $A(\vec R,\vec P)$, depending on position operator $\vec R$ and momentum operator $\vec P$, can be written as
\begin{align}\label{trHAppendix}
\trace\{A(\vec R,\vec P)\}=\int\frac{(\d\vec r)(\d\vec p)}{(2\pi\hbar)^D}\,A_W(\vec r,\vec p) ,
\end{align}
with the Wigner function
\begin{align}\label{WignerA}
A_W(\vec r,\vec p)&=\trace\{A(\vec R,\vec P) W(\vec R-\vec r,\vec P-\vec p)\}\nn\\
&=\int(\d\vec r')\,\mathrm{e}^{\frac{\I}{\hbar}\vec p\cdot\vec r'}\newbok{\vec r-\frac{\vec r'}{2}}{A(\vec R,\vec P)}{\vec r+\frac{\vec r'}{2}}%\\
%&=2^D\int(\d\vec q)\exp\left(-\frac{2\I}{\hbar}\vec q\cdot\vec r\right)\bok{\vec p-\vec q}{A(\vec R,\vec P)}{\vec p+\vec q}
\end{align}
of $A(\vec R,\vec P)$. $A_W(\vec r,\vec p)$ is also called Wigner transform, phase-space kernel, or Weyl symbol \cite{Zachos2005}, with \cite{Berge1989}
\begin{align}
W(\vec R-\vec r,\vec P-\vec p)=2^D\exp\left(-\frac{2\I}{\hbar}(\vec R-\vec r);(\vec P-\vec p)\right). \label{WignerKernel}
\end{align}
The semicolon in Eq.~(\ref{WignerKernel}) indicates ordering of products of the components $\vec R_i$ and $\vec P_i$, ${i=1,\dots,D}$, such that $\vec R_i$ always stands left of $\vec P_i$.

Approximations of $\trace\{A(\vec R,\vec P)\}$ can be introduced by approximating $A_W(\vec r,\vec p)$, followed by integrations over real numbers in Eq.~(\ref{trHAppendix}) instead of evaluating the trace in Hilbert space. For a product of arbitrary operators $A(\vec R,\vec P)$ and $B(\vec R,\vec P)$ one finds
\begin{align}\label{trAB2}
\trace\{ A\, B\}=\int\frac{(\d\vec r)(\d\vec p)}{(2\pi\hbar)^D}\,A_W(\vec r,\vec p)\,B_W(\vec r,\vec p) .
\end{align}
This relation is used in deriving Eq.~(\ref{density}). With the Moyal product or (Groenewold's) star product,
\begin{align}\label{ABW}
{[ A B]}_W=A_W\star B_W=A_W\,\mathrm{e}^{\I\frac{\hbar}{2}\mathrm{\Lambda}}B_W ,
\end{align}
see \cite{Groenewold1946,Moyal1949}, Eq.~(\ref{trAB2}) can be rewritten as
\begin{align}\label{trAB3}
\trace\{ A\, B\}=\int\frac{(\d\vec r)(\d\vec p)}{(2\pi\hbar)^D}\,A_W(\vec r,\vec p)\,\mathrm{e}^{\I\frac{\hbar}{2}\mathrm{\Lambda}}B_W(\vec r,\vec p) ,
\end{align}
where $\mathrm{\Lambda} $ is defined in Eq.~(\ref{Lambda}). That is, $A_WB_W$ and $A_W\exp\big(\I\frac{\hbar}{2}\mathrm{\Lambda} \big)B_W$ differ only by terms which integrate to zero over all phase space.

\renewcommand{\thesection}{Appendix \Alph{section}}
\section{\label{SemiclassicalexpansionofWignertransforms}Airy-averaged semiclassical expansions of Wigner transforms}
\renewcommand{\thesection}{\Alph{section}}

\setcounter{equation}{0}

In the following we give the derivation of Eqs.~(\ref{fAW}) and (\ref{trfA4}). We want to gain a systematic semiclassical expansion of the Wigner transform for arbitrary operator-valued functions $f( A)$, which cannot be calculated exactly for arbitrary potentials. The difficulty arises from the noncommutativity of the kinetic and the potential energy operator.

It suffices to consider the Wigner transform of the exponential since
\begin{align}\label{FourierfAppendix}
\big[f( A)\big]_W=\int\d\alpha \,g(\alpha)\,\big[\mathrm{e}^{\I\alpha  A}\big]_W ,
\end{align}
where $g(\alpha)$ are the Fourier components of $f(A)$. With the Airy function $\mathrm{Ai}(x)$ we define the `Airy average' of a function $f(x)$ as
\begin{align}
\Airy{f(x)}=\int\d x \,f(x)\, \mathrm{Ai}(x).
\end{align}
In the following we show that
\begin{align}
[\mathrm{e}^{\I\alpha A}]_W\cong\Airy{\mathrm{e}^{\I xt}}\left[1-\frac{\hbar^2}{16}\big\{A_W\mathrm{\Lambda}^2 A_W\big\}\partial_{A_W}^2\right]\mathrm{e}^{\I\alpha A_W}\label{107a}
\end{align}
is valid at $\mathcal O(\nab^2)$, indicated by `$\cong$', with $t$ given by
\begin{align}\label{tdef}
t^3=\frac18\alpha^3\hbar^2\big\{A_W\mathrm{\Lambda}  A_W\mathrm{\Lambda}  A_W\big\} .
\end{align}
The curly brackets $\{\,\}$ in Eqs.~(\ref{107a}) and (\ref{tdef}), as introduced in \cite{CinalBerge1993}, denote that the operators $\mathrm{\Lambda} $ only act inside the brackets and only on their neighboring functions $A_W$. For the Fourier transform of $\mathrm{Ai}(x)$, we get
\begin{align}
\Airy{\mathrm{e}^{\I xt}}=\mathrm{e}^{(\I t)^3/3}\cong 1-\I\alpha^3\frac{\hbar^2}{24}\big\{A_W\mathrm{\Lambda}  A_W\mathrm{\Lambda}  A_W\big\} .
\end{align}
Then, we may rewrite Eq.~(\ref{107a}) at $\mathcal O(\nab^2)$ as
\begin{align}
[\mathrm{e}^{\I\alpha  A}]_W&\cong \mathrm{e}^{\I\alpha A_W}\left[1-\frac{\hbar^2}{16}(\I\alpha)^2\big\{A_W\mathrm{\Lambda}^2 A_W\big\}+\frac{\hbar^2}{24}(\I\alpha)^3\big\{A_W\mathrm{\Lambda}  A_W\mathrm{\Lambda}  A_W\big\}\right] .\label{106}
\end{align}
With Eq.~(\ref{FourierfAppendix}), we obtain the well-known expression for $\big[f( A)\big]_W$ to $\mathcal O(\nab^2)$,
\begin{align}\label{fAWappendix}
\big[f( A)\big]_W&\cong f(A_W)-\frac{\hbar^2}{16}\big\{A_W\mathrm{\Lambda}^2 A_W\big\}f''(A_W)+\frac{\hbar^2}{24}\big\{A_W\mathrm{\Lambda}  A_W\mathrm{\Lambda}  A_W\big\}f'''(A_W) ,
\end{align}
thereby validating Eq.~(\ref{107a}). The gradient expansion of $\big[f( A)\big]_W$ has been addressed with numerous methods, also for orders higher than $\mathcal O(\nab^2)$; see for example \cite{Wigner1932,GrammaticosVoros1979,VonEiff1991}.

Eventually, we get $\big[f( A)\big]_W$ from Eqs.~(\ref{FourierfAppendix}) and (\ref{107a}),
\begin{align}\label{fAWAiry}
\big[f(A)\big]_W\cong\Airy{f\big(\tilde{A}_W\big)-\frac{\hbar^2}{16}\big\{A_W\mathrm{\Lambda}^2 A_W\big\}f''\big(\tilde{A}_W\big)},
\end{align}
viz.~Eq.~(\ref{fAWAiryMainText}), with $\tilde{A}_W$ given in Eq.~(\ref{fAWAiryMainTextAWtilde}). The approximate Wigner transform in Eq.~(\ref{fAWappendix}) is a special case of Eq.~(\ref{fAWAiry}). They coincide at order $\nab^2$. However, Eq.~(\ref{fAWAiry}) proves to be superior in the vicinity of the quantum-classical boundary, as argued in the present work.

The approximate Wigner transform in Eq.~(\ref{fAWAiry}) is applicable for any operator $A(\vec R,\vec P)$. We now derive Eq.~(\ref{trfA4}), a useful representation of the momentum integral
\begin{align}
\int(\d\vec p)\,\big[f( A)\big]_W(\vec r,\vec p)
\end{align}
for the special case in Eq.~(\ref{specialcase}), for which
\begin{align}\label{bwlambda0}
\big\{A_W\mathrm{\Lambda}  A_W\mathrm{\Lambda}  A_W\big\}=-\frac{(\vec p\cdot\nab)^2V(\vec r)}{m^2}-\frac{\big(\nab V(\vec r)\big)^2}{m}
\end{align}
and
\begin{align}
\big\{A_W\mathrm{\Lambda}^2 A_W\big\}=\frac{2}{m}\nab^2 V(\vec r) .\label{bwlambda1}
\end{align}
Furthermore, for any function $f$ that is isotropic in $\vec p$,
\begin{align}\label{bwlambda}
\int(\d\vec p)\,f(\vec p^2)\,\big\{A_W\mathrm{\Lambda}  A_W\mathrm{\Lambda}  A_W\big\}=\int(\d\vec p)\,f(\vec p^2)\left(-\frac{p^2\nab^2V(\vec r)}{D\,m^2}-\frac{\big(\nab V(\vec r)\big)^2}{m}\right) .
\end{align}
Inserting Eqs.~(\ref{bwlambda0}) and (\ref{bwlambda1}) into Eq.~(\ref{fAWappendix}), and using
\begin{align}
\int(\d\vec p)\,\vec p^2\,f'''\left(\frac{\vec p^2}{2m}\right)=-m\,D\int(\d\vec p)\,f''\left(\frac{\vec p^2}{2m}\right),
\end{align}
obtained from an integration by parts \footnote{We assume vanishing boundary contributions, $\big[p_j f''\big(\vec p^2/(2m)\big)\big]_{-\infty}^\infty=0$, for $j=1,\dots,D$.}, we get
\begin{align}\label{specialfAW}
\int(\d\vec p)\,\big[f(A)\big]_W(\vec r,\vec p)\cong \int(\d\vec p)\,\left[f(A_W)-\frac{\hbar^2}{12m}f''(A_W)\nab^2V-\frac{\hbar^2}{24m}f'''(A_W)(\nab V)^2\right].
\end{align}
%and
%\begin{align}\label{specialeAW}
%&\int(\d\vec p)\,\big[\mathrm{e}^{\I\alpha A}\big]_W(\vec r,\vec p)\cong \int(\d\vec p)\,\mathrm{e}^{\I\alpha A_W}\nn\\
%&\quad\times\Big[1-\frac{\hbar^2}{12m}(\I\alpha)^2\nab^2V-\frac{\hbar^2}{24m}(\I\alpha)^3(\nab V)^2\Big]\ ,
%\end{align}
%in accordance with Eq.~(\ref{FourierfAppendix}).
Analogously to the calculation presented for Eq.~(\ref{fAWappendix}), Eq.~(\ref{specialfAW}) can be given in terms of the Airy average, which leads to Eq.~(\ref{trfA4}). Along the same lines, we find
\begin{align}\label{specialeAWp2}
&\int(\d\vec p)\,\frac{\vec p^2}{2m}\,\big[f( A)\big]_W(\vec r,\vec p)\cong\int(\d\vec p)\frac{\vec p^2}{2m}\int\d x\, \mathrm{Ai}(x)\left[f\big(\tilde{A}_W\big)-\frac{\hbar^2(\nab^2V)}{12m}\frac{D-1}{D}f''\big(\tilde{A}_W\big)\right],
\end{align}
with ${\tilde{A}_W=\tilde{A}_W(\vec r,\vec p)=H_W(\vec r,\vec p)-\mu-x\, a(\vec r)}$. In the case of ${D=1}$ the gradient corrections enter only via $a(\vec r)$.

\section*{References}

\bibliography{/home/martintrappe/Desktop/PostDoc/PapersTalksPoster/myPostDocbib}

\end{document}